\documentclass[prd,aps,twocolumn,a4paper,floatfix,showpacs,nofootinbib,superscriptaddress]{revtex4-1}

\usepackage[utf8]{inputenc}
\usepackage[T1]{fontenc}
\usepackage{graphicx,psfrag}
\usepackage{mathrsfs}
\usepackage{amsmath,amsfonts,amssymb,pifont,gensymb}
\usepackage{multirow,enumerate}
\usepackage{comment,hyperref}
\usepackage{color}
\usepackage{acronym}
\usepackage{xspace}
\usepackage[normalem]{ulem}
\usepackage{mathtools}
\usepackage{subfigure}
\usepackage{makecell}

\newacro{BH}{black hole}
\newacro{NS}{neutron star}
\newacro{PN}{Post-Newtonian}
\newacro{BBH}{binary black hole}
\newacro{BNS}{binary neutron star}
\newacro{EOB}{effective-one-body}
\newacro{NR}{numerical relativity}
\newacro{GW}{gravitational wave}
\newacro{EOS}{equation-of-state}

\newcommand{\be}{\begin{equation}}
\newcommand{\ee}{\end{equation}}
\newcommand{\bea}{\begin{eqnarray}}
\newcommand{\eea}{\end{eqnarray}}
\newcommand{\bel}{\begin{align}}
\newcommand{\eel}{\end{align}}

\newcommand{\cmark}{\ding{51}}%
\newcommand{\xmark}{\ding{55}}%

\def\GMc2{{\rm G M_{\odot} c^{-2}}}

\def\mo{\hat{\omega}}

\def\NRtidal{\texttt{NRTidal}\xspace}
\def\NRTidal{\texttt{NRTidal}\xspace}
\def\NRTidalp{\texttt{NRTidalv2}\xspace}
\def\NRtidalp{\texttt{NRTidalv2}\xspace}

\def\PhenomPNRtidal{\texttt{IMRPhenomPv2\_NRTidal}\xspace}
\def\PhenomPNRtidal2{\texttt{IMRPhenomPv2\_NRTidal2}\xspace}

\def\SEOBNRv4T{\texttt{SEOBNRv4T}\xspace}

\usepackage{color}

\definecolor{cyan}{rgb}{0,0.9,0.9}
\definecolor{orange}{rgb}{0.9,0.5,0}
\definecolor{magenta}{rgb}{1,0,1}
\definecolor{purple}{rgb}{0.8,0.4,0.8}
\definecolor{gray}{rgb}{0.5,0.5,0.5}
\definecolor{mygreen}{rgb}{0.1,0.8,0.1}
\definecolor{darkblue}{rgb}{0.0,0.0,0.6}

\begin{document}

\title{Improving the \NRTidal model for binary neutron star systems}

\author{Tim \surname{Dietrich}}

\affiliation{Nikhef, Science Park, 1098 XG Amsterdam, The Netherlands}

\author{Anuradha Samajdar}

\affiliation{Nikhef, Science Park, 1098 XG Amsterdam, The Netherlands}

\author{Sebastian \surname{Khan}}

\affiliation{Max Planck  Institute for Gravitational Physics
(Albert Einstein Institute), Callinstr.~38, 30167 Hannover, Germany}

\affiliation{Leibniz Universit\"at Hannover, D-30167 Hannover, Germany}

\author{Nathan~K.~\surname{Johnson-McDaniel}}

\affiliation{DAMTP, Centre for Mathematical Sciences, Wilberforce Road, University of Cambridge, Cambridge, CB3 0WA, UK}

\author{Reetika Dudi}

\affiliation{Theoretical Physics Institute, University of Jena, 07743 Jena, Germany}

\author{Wolfgang \surname{Tichy}}

\affiliation{Department of Physics, Florida Atlantic University, Boca Raton, FL  33431, USA}

\date{\today}

\begin{abstract}
Accurate and fast gravitational waveform (GW) models are essential
to extract information about the properties of
compact binary systems that generate GWs.
Building on previous work,
we present an extension of the \NRTidal model
for binary neutron star (BNS) waveforms. \\
The upgrades are:
(i) a new closed-form expression for the tidal contribution to the
GW phase which includes further analytical knowledge and is calibrated
to more accurate numerical relativity data than previously available;
(ii) a tidal correction to the GW amplitude;
(iii) an extension of the spin-sector incorporating
equation-of-state-dependent finite size effects at quadrupolar and octupolar order; these appear
in the spin-spin tail terms and cubic-in-spin terms, both at 3.5PN. \\
We add the new description to the precessing binary black hole waveform model \texttt{IMRPhenomPv2} 
to obtain a frequency-domain precessing binary neutron star model. In addition, we extend the 
\texttt{SEOBNRv4\_ROM} and \texttt{IMRPhenomD} aligned-spin binary black hole waveform models with the improved 
tidal phase corrections.
Focusing on the new \texttt{IMRPhenomPv2\_NRTidalv2} approximant, 
we test the model by comparing with numerical relativity waveforms as well as
hybrid waveforms combining tidal effective-one-body and numerical relativity data.
We also check consistency against a tidal effective-one-body model across large regions of the BNS parameter
space. 
\end{abstract}

\maketitle

\section{Introduction}
\label{sec:intro}

The first gravitational wave (GW) signal
associated with electromagnetic (EM) counterparts,
detected on the 17th of August 2017, marks a breakthrough
in the field of multi-messenger astronomy~\cite{TheLIGOScientific:2017qsa,Monitor:2017mdv,GBM:2017lvd}.
Analyses of the GW and EM signatures favor a binary neutron star (BNS)
coalescence, e.g.,~\cite{Tanaka:2017qxj,
Nicholl:2017ahq,Tanvir:2017pws,Perego:2017wtu,Waxman:2017sqv,Metzger:2018uni,Abbott:2018wiz,
Kawaguchi:2018ptg,Hinderer:2018pei,Coughlin:2018fis,
Coughlin:2019kqf,Siegel:2019mlp}.
Due to the increasing sensitivity of advanced GW detectors,
multiple detections of merging BNSs are expected
in the near future~\cite{Aasi:2013wya}.

A prerequisite to extract information from the data
are theoretical predictions about the emitted GW signal.
The properties of the system are typically inferred via a coherent
Bayesian analysis based on cross-correlation of the measured
strain with predicted waveform approximants, e.g.,~\cite{Veitch:2014wba}.
These cross-correlations are done for a large number of target waveforms
and require large computational resources.
Thus, the computation of each individual waveform needs
to be efficient and fast to ensure that the Bayesian parameter estimation
of signals, containing several thousand GW cycles
(as typical for BNS systems), is at all manageable.
On the other hand, waveform models need to be accurate enough
to allow a correct estimate of the source properties, such as
the masses, the spins, and internal structure of the NSs.

Over the last years, there has been significant progress
modeling the GW signal associated with the BNS coalescence,
including the computation of higher-order tidal corrections or spin-tidal coupling,
e.g., Refs.~\cite{Damour:2012yf,Pani:2018inf,Banihashemi:2018xfb,
Abdelsalhin:2018reg,Landry:2018bil,Jimenez-Forteza:2018buh}, and
improved accuracy of BNS numerical relativity (NR)
simulations~\cite{Hotokezaka:2015xka,Bernuzzi:2016pie,Kiuchi:2017pte,
Dietrich:2018upm,Dietrich:2018phi,Foucart:2018lhe}.
However, although the analytical progress has improved
the performance of post-Newtonian (PN) waveform approximants,
PN models still become increasingly inaccurate towards the merger,
e.g.~\cite{Bernuzzi:2012ci,Favata:2013rwa,Wade:2014vqa,Hotokezaka:2016bzh,
Dietrich:2018uni,Dudi:2018jzn,Samajdar:2018dcx}.

Most of the current time-domain tidal waveform
models~\cite{Bernuzzi:2014owa, Hotokezaka:2015xka, Hinderer:2016eia, Steinhoff:2016rfi,
Nagar:2018zoe,Nagar:2018gnk,Akcay:2018yyh,Nagar:2018plt}
are based on the effective-one-body (EOB) description of the
general relativistic two-body problem~\cite{Buonanno:1998gg, Damour:2009wj}.
This approach has proven to be able to predict the BNS merger dynamics in large
regions of the BNS parameter space, but recent numerical relativity (NR) data
revealed configurations for which further improvements of the tidal EOB models are
required~\cite{Hotokezaka:2015xka, Dietrich:2017feu, Akcay:2018yyh}.
While one can expect that over the next years,
these issues will be overcome due to further progress in the fields of NR,
gravitational self-force, and PN theory,
the high computational cost for a single EOB waveform
is yet another disadvantage.
One possibility to speed up the EOB computation is the use of
high-order post-adiabatic approximations of the EOB
description to allow an accurate and efficient evaluation of
the waveform up to a few orbits before merger~\cite{Nagar:2018gnk}.
The other possibility, and most common approach,
is constructing reduced-order-models~\cite{Lackey:2016krb,Lackey:2018zvw}.
Those models allow the fast computation of waveforms in the frequency
domain and are well suited for a direct use in parameter
estimation pipelines.

In addition to PN and EOB approximants, there have been proposals
for alternative ways to describe tidal GW signals.
Refs.~\cite{Lackey:2013axa,Pannarale:2013uoa}
develop phenomenological black hole-neutron star
(BHNS) approximants based on NR data.
Ref.~\cite{Barkett:2015wia} transforms NR
simulations of binary black hole (BBH) systems by adding PN tidal effects,
and Refs.~\cite{Lange:2017wki,Lange:2018pyp} develop a method to employ
NR waveforms or computationally
expensive waveform approximants (such as tidal EOB waveforms)
directly for parameter estimation.

Another approach to describe BNS systems
was presented in Ref.~\cite{Dietrich:2017aum}, in which BBH models
have been augmented by an analytical closed-form expression correcting the
GW phase to include tidal effects.
This waveform model~\cite{Dietrich:2017aum,Dietrich:2018uni}, referred to as \NRTidal,
was implemented in the LSC Algorithm Library (LAL)~\cite{lalsuite} to support the analysis of
GW170817 by the LIGO and Virgo Collaborations
(LVC)~\cite{TheLIGOScientific:2017qsa,Abbott:2018wiz,Abbott:2018exr,
LIGOScientific:2018mvr,Abbott:2018lct} and has also been
used outside the LVC, e.g.~\cite{Dai:2018dca,Radice:2018ozg}.
In addition,
Ref.~\cite{Kawaguchi:2018gvj} developed an alternative tidal
approximant in the frequency domain combining
EOB and NR information following a similar idea 
as in Ref.~\cite{Dietrich:2017aum}.\\

Studies showed that for GW170817, with its signal-to-noise ratio (SNR) of $\sim30$,
waveform model systematics are within the statistical uncertainties,
i.e., that different employed tidal GW models give slightly
different, but consistent constraints on the binary properties,
e.g.,~\cite{Abbott:2018wiz}.
However, systematic effects will grow for an increasing number of
detections or GW observations with larger
SNRs~\cite{Dudi:2018jzn,Samajdar:2018dcx}.
Ref.~\cite{Samajdar:2018dcx} stated that for a
GW170817-like event measured with the anticipated design sensitivity of the Advanced LIGO
and Advanced Virgo detectors, systematic effects will dominate and
the extracted equation of state (EOS) constraints between existing waveform
approximants will become inconsistent.
Furthermore, the analysis presented in Ref.~\cite{Dietrich:2018uni,Dudi:2018jzn}
showed that the original \texttt{NRTidal} model could potentially underestimate
tidal deformabilities, leading to possible biases for future detections with larger SNRs.

Therefore, to further push for the availability of a fast and accurate waveform model
employable for the upcoming observing runs in the advanced detector era, after recalling the
basics of \NRTidal and discussing the NR simulations and hybrid waveform construction in Sec.~\ref{sec:previous}, we improve
the \NRTidal description by:
\begin{enumerate}[(i)]
 \item Recalibrating the closed-form phenomenological tidal description including
additional analytical knowledge and using improved NR data (Sec.~\ref{sec:improvements:recalibration});
 \item Adding a tidal GW amplitude correction to the model (Sec.~\ref{sec:improvements:tidal_ampl});
 \item Incorporating EOS-dependent 3.5PN spin-spin and cubic-in-spin effects proportional to the
 quadrupole and octupole moments of the NSs~\cite{Marsat:2014xea,Bohe:2015ana,Krishnendu:2017shb,Nagar:2018plt} 
 (Sec.~\ref{sec:improvements:spin}).
\end{enumerate}
We validate the new \NRTidalp approximant with a set of $10$ high-resolution NR waveforms (Sec.~\ref{sec:validation:NR}) 
and $18$ hybrids of NR waveforms and the \texttt{TEOBResumS} tidal EOB model~\cite{Nagar:2018zoe} 
(Sec.~\ref{sec:validation:hybrids}).
Furthermore, we compare the model in a larger region of the parameter space than currently covered
with NR simulations by computing the mismatch with respect to
the \texttt{SEOBNRv4T} tidal EOB model~\cite{Hinderer:2016eia,Steinhoff:2016rfi} (Sec.~\ref{sec:validation:mismatches}).
We note that for this waveform model Ref.~\cite{Lackey:2018zvw} recently developed a reduced order model 
which can also be used directly for GW data analysis. 
We conclude in Sec.~\ref{sec:summary}. In the Appendices, we discuss possible extensions to the model, 
considering the tidal amplitude correction 
(Appendix~\ref{app:tidal_amp}) and the mass ratio dependence of the tidal phase (Appendix~\ref{app:mass_ratio}).\\

In this article geometric units are used by setting $G=c=M_\odot=1$. At
some places units are given explicitly to allow a better interpretation.
Further notations are $M=M_A + M_B$
for the total mass of the system, $\chi_A,\chi_B,\Lambda_A,\Lambda_B$
for the individual dimensionless spins and tidal deformabilities
of the stars. The mass ratio of the system is $q=M_A/M_B$ and
the symmetric mass ratio is $\nu=M_A M_B/(M_A+M_B)^2$.
We define the labeling of the individual
stars so their masses satisfy $M_A\geq M_B$.

\section{Basic ideas and improved numerical relativity data}
\label{sec:previous}

\subsection{The basic idea of NRTidal}

During the BNS coalescence, 
each star gets deformed due to the gravitational field of the companion. 
These tidal deformations accelerate the inspiral 
and leave a clear imprint in the GW 
signal, e.g.,~\cite{Flanagan:2007ix}. 
Consequently, the theoretical modeling of BNSs and the extraction of tidal effects 
from measured GW signals is an important way of determining the internal 
structure of NSs and thus the EOS of supranuclear dense matter.

The complex time-domain GW signal is given by 
\begin{equation}
h(t)=A(t)e^{-{\rm i}\phi(t)}, \label{eq:h_TD}
\end{equation}
with amplitude $A(t)$ and
time-domain phase $\phi(t)$. Here we only consider the dominant $2,2$ [spin $(-2)$ weighted spherical harmonic] mode.
We assume in the following that the phase can be decomposed into
\begin{equation}\label{eq:phi_omg}
 \phi(\mo) = \phi_{\rm pp} (\mo) + \phi_{\rm SO}(\mo) + \phi_{\rm SS}(\mo) + \phi_{\rm T}(\mo) + \cdots \ ,
\end{equation}
where the dimensionless GW frequency is given by
$\mo = M\omega=M\partial_t\phi(t)$.
Here $\phi_{\rm pp}$ denotes the nonspinning, point-particle,
contribution to the overall phase, $\phi_{\rm SO}$ corresponds
to contributions caused by spin-orbit coupling,
$\phi_{\rm SS}$ corresponds to contributions caused by spin-spin effects
(both self-spin and spin-interactions),
and $\phi_{\rm T}$ denotes the tidal effects present in the GW phase.

Similar to Eq.~\eqref{eq:h_TD}, the waveform
can be written in the frequency domain as
\begin{equation}
\tilde{h}(f)=\tilde{A}(f)e^{-{\rm i}\psi(f)}, \label{eq:h_FD},
\end{equation}
with GW frequency $f$ and frequency domain amplitude $\tilde{A}(f)$ and phase $\psi(f)$.
Here we assume again:
\begin{equation}\label{eq:psi_omg}
 \psi(\mo) = \psi_{\rm pp} (\mo) + \psi_{\rm SO}(\mo) + \psi_{\rm SS}(\mo) + \psi_{\rm T}(\mo)+ \cdots \ .
\end{equation}

Constraints on the supranuclear EOS governing the
matter inside NSs rely on an accurate measurement of the tidal phase contribution.
This contribution enters first at the 5th PN order.\footnote{
There is also the possibility of extracting EOS information
from the spin-spin interaction first entering in the 2PN $\psi_\text{SS}$ contribution,
where the individual terms of $\psi_\text{SS}$ are proportional to the square of the
individual spins, i.e., $\chi_A^2$, $\chi_B^2$, or $\chi_A \chi_B$.
Although the maximum NS spin in a BNS is not precisely known,
the fastest spinning NS in a BNS system observed to date
(PSR J1946+2052~\cite{Stovall:2018ouw})
will only have a dimensionless spin of $\sim 0.02$--$0.04$ at merger~\cite{Kumar:2019xgp}.
Thus, obtaining EOS information from the spin-spin phase contribution is extremely challenging.
}

The main idea of the \NRtidal approach is to provide a closed-form approximation
for the tidal phase $\phi_{\rm T}$ or $\psi_{\rm T}$.
Because standard GW data analysis is carried out in the frequency domain, the frequency domain model
is of particular importance, due to its efficiency.
In addition to the tidal contribution, the final \NRTidal approximant also
incorporates EOS dependent effects in $\psi_{\rm SS}$,
since the spin-spin contributions depend on the quadrupole and
higher moments of the individual stars, and thus on the internal structure of the stars.

We note that there are higher-order spin-tidal coupling effects that have recently been computed~\cite{Abdelsalhin:2018reg, Landry:2018bil}.
However, as outlined in~\cite{Jimenez-Forteza:2018buh}, these terms will be unmeasurable
in the advanced GW detector era. Therefore, we do not include them in the current
description to avoid unnecessary computational costs.

\subsection{High-precision NR simulations}
\label{sec:NR}

\begin{figure}[t]
\includegraphics[width=0.5\textwidth]{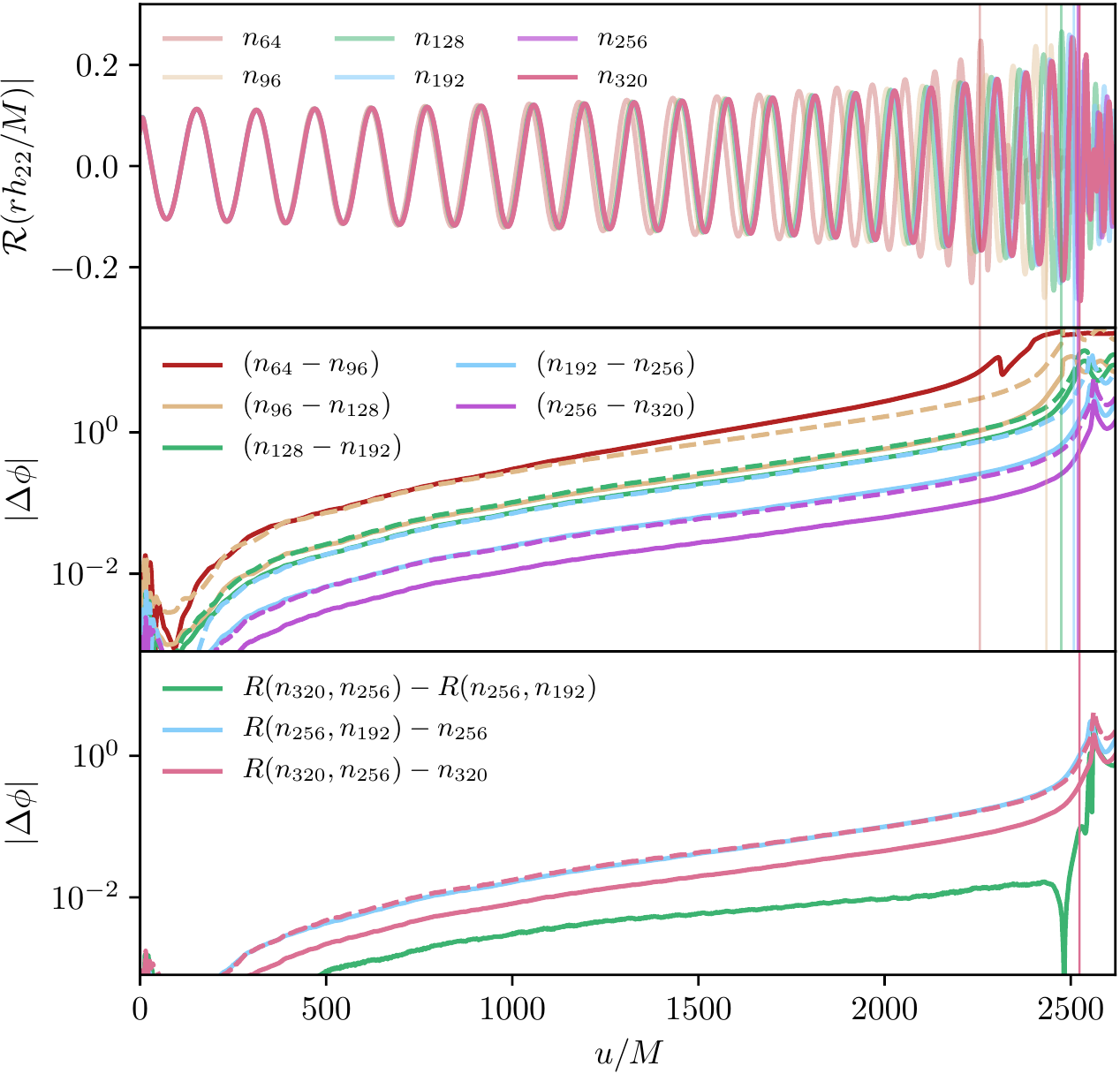}
\caption{New high-resolution NR data employed for the calibration
of the \NRTidalp approximant.
Top panel: Real part of the GW signal for the six different resolutions
employing $64$, $96$, $128$, $192$, $256$, and $320$ points in the refinement levels covering
the individual NSs.
The waveforms shown are already extrapolated to spatial infinity to correct for the finite radius extraction;
see~\cite{Bernuzzi:2016pie} for more details (we use $K = 1$ here).
Middle panel:
Phase difference between different resolutions.
Bottom panel:
Phase difference between different Richardson extrapolated
waveforms or between a Richardson extrapolated
waveforms and the waveform from an individual resolution.
The vertical lines in each panel refer to the time of merger, i.e.,
the peak time of the GW amplitude for the individual resolutions.
The dashed lines in the bottom two panels show the phase difference scaled to the next lowest pair of resolutions
assuming second order convergence. $u$ denotes the retarded time.}
\label{fig:NR_data}
\end{figure}

The field of NR has made significant progress
over the last years. Nevertheless, the production of highly
accurate gravitational BNS waveforms remains challenging and
there exist only a small number of simulations with low eccentricity and
with phase errors small enough to allow GW modeling;
cf.~Refs.~\cite{Hotokezaka:2015xka,Bernuzzi:2016pie,
Dietrich:2018upm,Dietrich:2018phi,Kiuchi:2017pte,Foucart:2018lhe}.

In addition to the dataset used for the original \texttt{NRTidal}
calibration~\cite{Dietrich:2017aum,Dietrich:2018uni}, we performed one
additional simulation for a non-spinning equal-mass BNS setup
employing a piecewise-polytropic parametrization of the SLy~\cite{Read:2008iy} EOS.
This EOS is in agreement with recent constraints extracted
from GW170817~\cite{Abbott:2018exr,LIGOScientific:2018mvr,Bauswein:2017vtn,
Most:2018hfd,Radice:2018ozg,Coughlin:2018fis} and thus
is a natural choice for our work.\footnote{While the maximum mass of $2.05 M_\odot$ of the
SLy EOS is slightly outside of the $68.3\%$ credible region of the recent heavy pulsar mass measurement
in~\cite{Cromartie:2019kug} ($[2.07,2.28] M_\odot$), it is well inside the $95.4\%$ credible region of $[1.97,2.40]M_\odot$, which is
why we still consider it here.}
The same physical configuration has already been used in the
past for the construction of the \NRTidal model~\cite{Dietrich:2017aum,Dietrich:2018upm};
cf.~Table~\ref{tab:calibrating_hybrids} for further details.
In~\cite{Dietrich:2017aum,Dietrich:2018upm}, 
we have simulated this setup with the BAM code~\cite{Bruegmann:2006at,Thierfelder:2011yi,
Dietrich:2015iva,Bernuzzi:2016pie} for $5$ different resolutions with
$64$, $96$, $128$, $192$, and $256$ points in the finest refinement level covering the individual NSs.
Here, we add one additional simulation with $320$ points in the finest refinement level.
This corresponds to a spatial resolution of $0.047 M_\odot \approx  70 \ \rm m$ and
computational costs of $\sim 5$~million CPU-hours for this single resolution.

The availability of six different resolutions and the presence of clean convergence across
multiple resolutions allows us to employ Richardson extrapolation to
obtain an improved GW signal and to provide an associated error 
budget; see Ref.~\cite{Bernuzzi:2016pie} for more details.
We present the GW signal for the different resolutions
in Fig.~\ref{fig:NR_data} (top panel) and the phase difference
and convergence properties in the middle and bottom panels.

Except for the lowest resolution, clean second order convergence 
is obtained throughout the inspiral.
This becomes evident by comparison of the individual phase 
differences with the phase differences rescaled 
assuming second order convergence (dashed lines).
For the lowest resolution setup ($n_{64}$), second order convergence is lost
a few orbits before merger ($u \approx 1500 M$).
Merger times for each resolution are indicated by 
vertical solid lines in Fig.~\ref{fig:NR_data}.

The phase difference between the highest ($n_{320}$) and second highest resolution ($n_{256}$) is
$0.52\ \rm rad$ at the moment of merger.
Performing a Richardson extrapolation~\cite{Bernuzzi:2016pie},
we obtain more accurate phase descriptions. We denote the Richardson extrapolated
data obtained from the resolutions $n_X$ and $n_Y$ as $R(n_{X},n_{Y})$.
We crosscheck the robustness of the procedure by presenting
the phase differences $R(n_{320},n_{256})-n_{320}$ and $R(n_{256},n_{192})-n_{256}$
in the bottom panel of Fig.~\ref{fig:NR_data}. Rescaling the phase difference
of $R(n_{320},n_{256})-n_{320}$ assuming second order convergence shows
excellent agreement with $R(n_{256},n_{192})-n_{256}$. This demonstrates that the 
leading error term scales quadratically with respect to the grid spacing/resolution. 

Thus, we can estimate the uncertainty of the Richardson extrapolated waveform
$R(n_{320},n_{256})$ to be the difference with the $n_{320}$ resolution. At the moment of merger,
this gives an uncertainty of $0.37\ \rm rad$. At this time the estimated error due to the finite radius extraction
is below $0.044\ \rm rad$, which leads to a conservatively estimated total error 
of $\Delta \phi_{\rm mrg} \lesssim 0.38\ \rm rad $ at merger.

An alternative, but not conservative, error measure is given
by the difference between the two Richardson extrapolated waveforms (green line in the bottom panel).
We find that throughout the inspiral the difference between the $R(n_{320},n_{256})$ and $R(n_{256},n_{192})$
is below $0.1\ \rm rad$ (at the moment of merger $\Delta \phi = 0.087\ \rm rad$, which would lead to a total
error of $\lesssim 0.1\ \rm rad$ once finite radius extraction is included).

In addition to this new setup, we also consider the additional two high resolution simulations 
available in the CoRe database~\cite{Dietrich:2018phi}, cf.~Table~\ref{tab:calibrating_hybrids}. 
These setups, CoRe:BAM:0037 and CoRe:BAM:0064, only employ 192 points across the star and 
have conservatively estimated phase uncertainties at merger of $1.20\ \rm rad$ and $2.27\ \rm rad$, 
respectively. We incorporated this accuracy difference by weighting the individual setups differently 
during the construction of the \NRTidalp phase, as discussed in the next subsection. 

\subsection{Hybrid Construction}
\label{sec:hybrid}

\begin{table}[t]
  \centering
  \caption{The non-spinning BNS and BBH hybrids employed in the construction of the \NRtidalp model.
           The columns refer to the name, the employed EOS, the individual masses of the stars $M^A$, $M^B$,
           the tidal deformabilities $\Lambda^A,\Lambda^B$, the tidal coupling constant $\kappa^T_{\rm eff}$ [Eq.~\eqref{eq:kappa}],
           and the ID in the CoRe and SXS databases.}
\begin{footnotesize}
\begin{tabular}{l||cccccccccccc}
\hline
  Name & EOS & $M_{A}$ & $M_B$& $\Lambda_A$ & $\Lambda_B$ & $\kappa^T_{\rm eff}$ & ID \\
\hline
SLy  & SLy  & 1.350 & 1.350 & 392.1  & 392.1 & 73.5   & CoRe:BAM:0095\footnote{Our work employs a higher resolution
than currently available for this setup in the CoRe catalog.} \\
H4   & H4   & 1.372 & 1.372 & 1013.4 & 1013.4 & 190.0 & CoRe:BAM:0037 \\
MS1b & MS1b & 1.350 & 1.350 & 1389.4 & 1389.4 & 288.1 & CoRe:BAM:0064 \\
BBH  & --   & 1.350 & 1.350 & 0     & 0     & 0     & SXS:BBH:0066 \\
\hline
\end{tabular}
\end{footnotesize}
\label{tab:calibrating_hybrids}
\end{table}

In the original \texttt{NRTidal} work,
PN, EOB, and NR approximants have been separately used in
different frequency intervals. Here, we start by constructing
hybrid waveforms consisting of a time domain
tidal EOB model (\texttt{TEOBResumS}) inspiral~\cite{Nagar:2018zoe}
connected to the high resolution NR simulation discussed above.
The hybridization is performed as discussed in
Refs.~\cite{Dietrich:2018uni,Dudi:2018jzn} to which we refer for further details.
In addition to the BNS hybrid waveforms, we also create a hybrid between the
non-tidal version of the \texttt{TEOBResumS} model and a binary black hole waveform
computed with the SpEC code~\cite{SpEC}, setup
\texttt{SXS:BBH:0066} of the public SXS catalog~\cite{Mroue:2013xna,Blackman:2015pia}.
All the hybrids have an initial frequency of $20\ \rm Hz$.

\begin{figure}[t]
\includegraphics[width=0.5\textwidth]{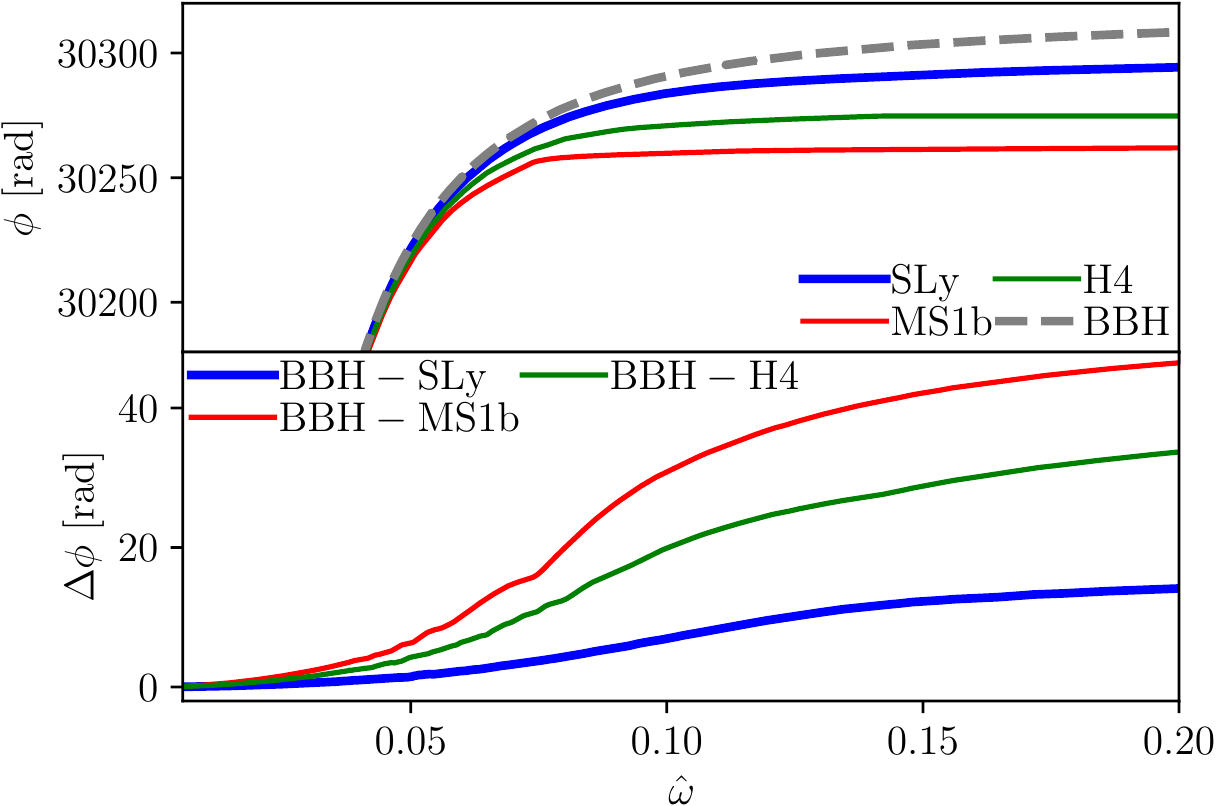}
\caption{Time domain phase of the hybrid waveforms employed to develop \NRTidalp.
The bottom panel shows the phase difference caused by tidal effects.}
\label{fig:hybrid}
\end{figure}

We present the time domain phase evolution of the BBH and BNS hybrids
in Fig.~\ref{fig:hybrid}.
For this plot we align the waveforms at $\sim 22\ \rm Hz$.

We emphasize that only the four hybrid waveforms listed
in Table~\ref{tab:calibrating_hybrids} are used for calibration of the \NRTidalp model,
where the dataset we are going to fit is 
\begin{small}
\begin{equation}
\begin{split}
 \phi_T^{\rm NR} &= \frac{1}{320+192+192} \bigg[ 320 (\phi_{\rm SLy}- \phi_{\rm BBH})  +
                     192(\phi_{\rm H4}- \phi_{\rm BBH})   \\ 
                 &  \quad +  192 (\phi_{\rm MS1b}- \phi_{\rm BBH})  \bigg] \\ 
                 & =  \frac{1}{11} \bigg( 5 \phi_{\rm SLy} + 3 \phi_{\rm H4} + 3 \phi_{\rm MS1b}
                     - 11  \phi_{\rm BBH} \bigg). \label{eq:NRdata_construction}
\end{split}
\end{equation}
\end{small}
The factors are obtained by linearly weighting the resolutions of the individual NR data, i.e., 
320 points accross the star for the SLy setup and 192 for H4 and MS1b setups. 
We decided to use this minimal dataset since
these are the available data with the highest accuracy. Note that  
a simple restriction to the highest resolution, i.e., the SLy data, leads to a 
phase description which does not accurately characterize binaries with large
tidal deformabilities. 
Thus, it would be preferable to include in the future a larger number
of NR simulations with varying masses, spins, mass ratios, and EOSs once these are available.
However, while there are a small number of
high quality waveforms~\cite{Dietrich:2018upm}, these waveforms do
not span a sufficiently large region of the parameter space to incorporate
additional mass ratio, EOS, or mass dependencies 
in our phenomenological ansatz.
\section{Improvements}
\label{sec:improvements}

The \NRTidalp approach can be added to any BBH model:
we focus our discussion here on the frequency-domain \texttt{IMRPhenomPv2}, \texttt{IMRPhenomD}, 
and \texttt{SEOBNRv4\_ROM} models.
We primarily concentrate on the extension of 
\texttt{IMRPhenomPv2}~\cite{Hannam:2013oca, Khan:2015jqa} describing precessing systems.
In addition, we have also added the improved tidal phase description to the 
\texttt{SEOBNRv4\_ROM}~\cite{Bohe:2016gbl} and the 
\texttt{IMRPhenomD}~\cite{Khan:2015jqa} approximants.\footnote{See~\cite{Purrer:2014fza,Purrer:2015tud} for more details of the reduced order model
technique used to construct \texttt{SEOBNRv4\_ROM} from the time domain approximant \texttt{SEOBNRv4}.} 
For \texttt{SEOBNRv4\_ROM} and \texttt{IMRPhenomD}, we decided to include 
only the tidal phase description to reduce additional computational 
costs and allow a faster computation of waveforms 
than for \texttt{IMRPhenomPv2\_NRTidalv2}. 

We present an overview of all existing \NRTidal models 
in Table~\ref{tab:models}. 

\begin{table*}
  \centering
  \caption{Overview of the existing \NRTidal approximants.
  The individual columns refer to: the name of the approximant, 
  the BBH baseline, the employed tidal phase, 
  the employed spin-spin and cubic-in-spin contribution, 
  employed tidal amplitude corrections, and the incorporation of precession, 
  as well as the computational time $\Delta T_{\rm f_{min}}$ 
  of the model to produce a single waveform for a non-spinning, equal mass binary with 
  individual masses $M_{A,B}=1.35$ and $\Lambda_{A,B}=400$ 
  on an Intel Xeon E5-2630v3 processor for various starting frequencies.}
\begin{footnotesize}
\begin{tabular}{l||c|c|c|c|c|c||cccc}
\hline
\hline
\multirow{2}{*}{LAL approximant name}               & \multirow{2}{*}{BBH baseline}                & \multirow{2}{*}{$\psi_T$} & \multirow{2}{*}{spin-spin} & \multirow{2}{*}{cubic-in-spin} & \multirow{2}{*}{tidal amp.} & \multirow{2}{*}{precession} 
& \multicolumn{4}{c}{$\Delta T_{\rm f_{min}}\ \rm [s]$} \\
& &  &&  & & & $10 \rm \ Hz$ & $20\rm \ Hz$ & $30\rm \ Hz$ & $40\rm \ Hz$ \\
\hline
\texttt{IMRPhenomD\_NRTidal}      & \texttt{IMRPhenomD}     & \NRTidal       & up to 3PN (BBH)      & \xmark      & \xmark & \xmark & 
 2.55 & 0.29 & 0.14 & 0.07 \\
\texttt{IMRPhenomD\_NRTidalv2}    & \texttt{IMRPhenomD}     & \NRTidalp      & up to 3PN            & \xmark      & \xmark & \xmark & 
 2.54 & 0.29 & 0.14 & 0.07 \\ 
\texttt{SEOBNRv4\_ROM\_NRTidal}   & \texttt{SEOBNRv4\_ROM}  & \NRTidal       & up to 3PN            & \xmark      & \xmark & \xmark & 
 3.39 & 0.40 & 0.18 & 0.09 \\
\texttt{SEOBNRv4\_ROM\_NRTidalv2} & \texttt{SEOBNRv4\_ROM}  & \NRTidalp      & up to 3PN            & \xmark      & \xmark & \xmark & 
 3.34 & 0.40 & 0.18 & 0.09 \\
\texttt{IMRPhenomPv2\_NRTidal}    & \texttt{IMRPhenomPv2}   & \NRTidal       & up to 3PN            & \xmark      & \xmark & \cmark & 
 7.30 & 0.90 & 0.43 & 0.21 \\
\texttt{IMRPhenomPv2\_NRTidalv2}  & \texttt{IMRPhenomPv2}   & \NRTidalp      & up to 3.5PN          & up to 3.5PN & \cmark & \cmark & 
 8.56 & 1.06 & 0.51 & 0.28 \\
\hline
\end{tabular}
\end{footnotesize}
 \label{tab:models}
\end{table*}

\subsection{Recalibrating the NRTidal phase}
\label{sec:improvements:recalibration}

\begin{figure}[t]
\includegraphics[width=0.5\textwidth]{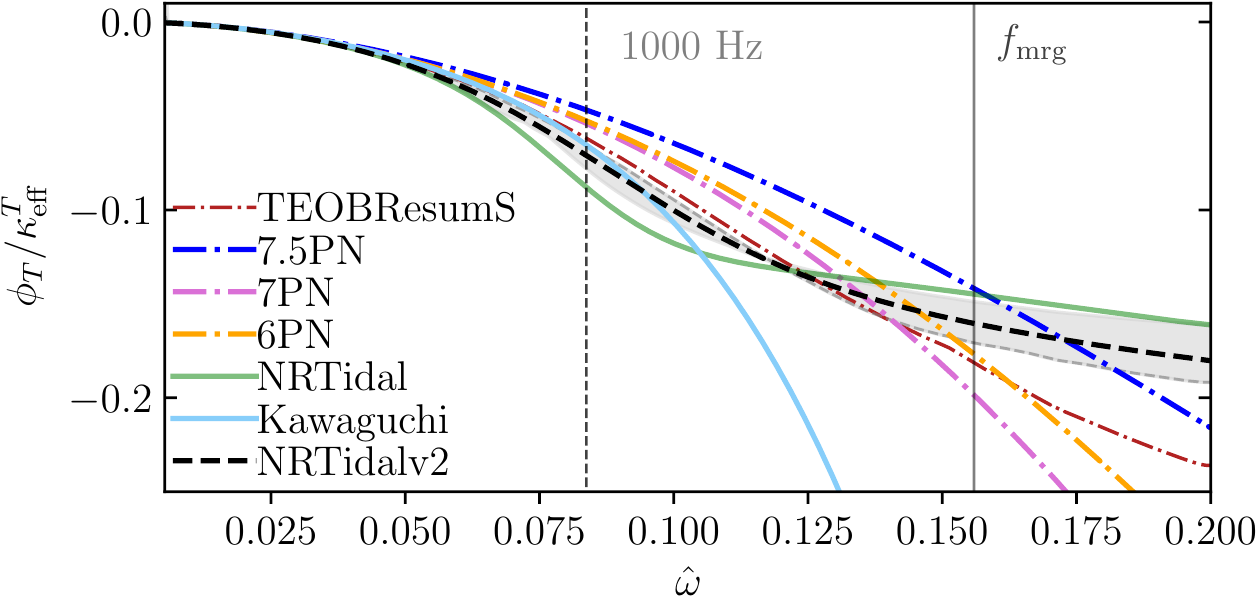}
\caption{Time domain tidal phase contribution divided by the tidal 
coupling constant $\kappa^T_{\rm eff}$. 
We show as a gray shaded region the parameter space covered by our 
hybrids (Table~\ref{tab:calibrating_hybrids}) and as a dashed gray line the SLy hybrid's
tidal phase divided by the coupling constant for this setup. 
In addition, the 6PN (orange), 7PN (orchid), and 7.5PN (blue) tidal phase estimates,
as well as the original \texttt{NRTidal}~\cite{Dietrich:2017aum} (green)
and Kawaguchi~\emph{et al.}~\cite{Kawaguchi:2018gvj} (cyan) approximants are presented. 
As a dark red line, we present an estimate obtained using the tidal EOB model \texttt{TEOBResumS}. 
We note that the Kawaguchi~\emph{et al.}\ and \texttt{TEOBResumS} tidal phases depend  
on $\tilde{\Lambda}$ even after being scaled by the coupling constant; 
for this plot and the following ones we use 
$\tilde{\Lambda}= 392.1$ to describe the SLy configuration. 
\NRTidalp is shown with a black dashed line. 
We mark the frequencies of $1000\ \rm Hz$ and the merger frequency corresponding 
to our SLy setup, described in Table~\ref{tab:calibrating_hybrids}.}
\label{fig:fit_td}
\end{figure}

\subsubsection{Ansatz for the NRTidal time-domain phase}

Non-spinning tidal contributions start entering the GW phasing at
the 5PN order and partially known analytical knowledge exists up
to 7.5PN~\cite{Damour:2012yf}:
\begin{align}
\phi_T & = - \kappa_A c^A_{\rm Newt} x^{5/2}
\left(1 + c^A_1 x + c^A_{3/2} x^{3/2}+ \right. \nonumber \\
& \quad + \left.  c^A_2 x^2 + c_{5/2}^A x^{5/2} \right) + [ A \leftrightarrow B ], \label{eq:PhiT_DNV}
\end{align}
with the dimensionless EOB tidal parameter $\kappa_A$ (defined below) and
$x(\mo)=(\mo/2)^{2/3}$.
The individual coefficients $c_i^A$ are
\begin{small}
\begin{subequations}
\begin{eqnarray}
 c^A_{\rm Newt}  &  = &  -\frac{(X_A+12 X_B)}{8 X_A X_B^2}, \\
 c^A_1 & = & -\frac{5 \left(260 X_A^3-2286 X_A^2-919 X_A+3179\right)}{336 \left(11 X_A-12\right)} , \\
 c^A_{3/2} & = & -\frac{5}{2} \pi , \\
 c_2^A & = & \left[ 5 \left(67702048 X_A^5-223216640 X_A^4+337457524 X_A^3 \right . \right.  \nonumber \\
            &&  \left. \left. -141992280 X_A^2+96008669 X_A-143740242\right) \right]/ \nonumber \\
            &&  \left[ 3048192 \left(11 X_A-12\right) \right] , \\
 c_{5/2}^A & = & -\frac{\pi  \left(10232 X_A^3-7022 X_A^2+22127 X_A-27719\right)}{192 \left(11 X_A-12\right)} , \label{eq:PN_coeff}
\end{eqnarray}
\end{subequations}
\end{small}
and similarly with $A \leftrightarrow B$. Here $X_{A,B}=M_{A,B}/M$.
We note that although analytic knowledge exists up to the 7.5PN order,
some unknown terms are present at 7PN. 
As discussed in Ref.~\cite{Damour:2012yf}, 
these terms are expected to be small and are set to zero in our definition of
$c_2^A$, $c_2^B$.

As in the original 
\texttt{NRTidal} description~\cite{Dietrich:2017aum,Dietrich:2018upm}
we introduce the effective tidal coupling constant $\kappa_{\rm eff}^T$
which describes the dominant tidal and mass ratio effects:
\begin{equation}
\label{eq:kappa}
\kappa^T_{\rm eff} = \frac{2}{13} \left[
\left(1+12\frac{X_B}{X_A}\right)\left(\frac{X_A}{C_A}\right)^5 k^A_2 +
 (A \leftrightarrow B) \right]  \ ,
\end{equation}
where $C_{A,B}\equiv M_{A,B}/R_{A,B}$ are the compactnesses of the stars at isolation,
and $k^{A,B}_2$ the Love numbers describing the static quadrupolar deformation
of one body in the gravitoelectric field of the 
companion~\cite{Damour:1983LesHouches,Hinderer:2007mb,Damour:2009vw,Binnington:2009bb}.
The parameter $\kappa^T_{\rm eff}$ is related to $\tilde{\Lambda}$
(the mass-weighted tidal deformability
commonly used in GW analysis~\cite{Wade:2014vqa}) by
\begin{equation}
\tilde{\Lambda}=\frac{16}{3} \kappa^T_{\rm eff},
\end{equation}
and the individual tidal deformability parameters are given by
\begin{equation}
\Lambda_{A,B}=\frac{2}{3} \frac{k_2^{A,B}}{C_{A,B}^5}. 
\end{equation}
The EOB tidal parameter used in Eq.~\eqref{eq:PhiT_DNV} is given by
$\kappa_A = 3X_BX_A^4\Lambda_A$.

In the following, we restrict the parameters $c_1^{A,B},c_{3/2}^{A,B},c_2^{A,B},c_{5/2}^{A,B}$
to their equal-mass values (due to the absence of a large set of high-quality
unequal mass NR data),
and therefore, discard the superscripts $A$ and $B$.
For this case, an effective representation of tidal effects is obtained using
\begin{equation}
\phi_T (x) = - \kappa_{\rm eff}^T
\frac{13}{ 8 \nu} x^{5/2} P_{\rm NRTidalv2}(x) \ ,  \label{eq:NRTidal_TD}
\end{equation}
with the Pad\'e approximant
\begin{small}
\begin{equation}
P_{\rm NRTidalv2}(x) = \frac{1 + n_1 x+ n_{3/2} x^{3/2} + n_2 x^2 + n_{5/2} x^{5/2} + n_3 x^3}
{1+ d_1 x+ d_{3/2} x^{3/2}+ d_2 x^2}. \label{eq:Pade_TD}
\end{equation}
\end{small}
To enforce consistency with the analytic PN knowledge [Eqs.~\eqref{eq:PhiT_DNV}-\eqref{eq:PN_coeff}],
some of the individual terms are restricted
\begin{small}
\begin{subequations}
\begin{eqnarray}
 n_1     & = & c_1 + d_1, \\
 n_{3/2}     & = & \frac{c_1 c_{3/2}-c_{5/2}-c_{3/2} d_1 + n_{5/2}}{c_1}, \\
 n_{2} & = & c_2 + c_1 d_1 + d_2 , \\
 d_{3/2} & = & - \frac{c_{5/2}+c_{3/2} d_1 - n_{5/2}}{c_1},
 \end{eqnarray}
  \end{subequations}
\end{small}
with
\begin{small}
\begin{subequations}
\begin{align}
  c_{1}   & =  \frac{3115}{624}, \qquad
  & c_{3/2} & = -\frac{5 \pi}{2}, \\
  c_2     & =  \frac{28024205}{1100736},  \qquad
  & c_{5/2} & =  -\frac{4283 \pi}{312}.
      \end{align}
      \end{subequations}
      \end{small}
The remaining unknown $4$ parameters are fitted to the data:
\begin{small}
\begin{subequations}
\begin{align}
  n_{5/2} & =  312.48173, \qquad
  & n_3   & = -342.15498, \\
  d_1     & = -20.237200,  \qquad
  & d_2  & =  -5.361630.
      \end{align}
      \end{subequations}
\end{small}

Figure~\ref{fig:fit_td} shows our findings.
We show as a gray 
shaded region the parameter space in $\phi_T/\kappa_{\rm eff}^T$ covered 
by our simulations, where the gray dashed line refers explicitly to the 
SLy configuration. Here we do not include any error estimate in the BBH 
hybrid used to extract the tidal phase.
In addition, we present the 6PN tidal contribution, 
which the old \texttt{NRTidal} approximant reduces to in the low frequency limit;
the 7.5PN contribution, which the new \NRTidalp reduces to in the low frequency limit;
and the 7PN contribution, which is the PN approximant showing the best agreement to the NR data.
We also show the tidal phase
given in Kawaguchi~\emph{et al.}~\cite{Kawaguchi:2018gvj},\footnote{We obtain the time domain
tidal phase approximant from the frequency domain expression given in
Ref.~\cite{Kawaguchi:2018gvj} using the stationary phase approximation.}
which has been calibrated to NR simulations up to a frequency
of $1000 \ \rm Hz$ (thin dashed line). The model of Ref.~\cite{Kawaguchi:2018gvj}
loses validity outside its calibration region and overestimates 
tidal effects at the moment of merger, though this would not affect GW data analysis if a maximum 
frequency of $1000 \ \rm Hz$ is employed, or the signal at frequencies $\gtrsim 1000 \ \rm Hz$ is sufficiently suppressed by the detectors' noise. 
In addition, we find good agreement between the Kawaguchi~\emph{et al.}~fit
and the new \NRTidalp approximation below $1000 \ \rm Hz$.
We also show the estimated tidal phase extracted by comparing our BBH hybrid with
a tidal EOB waveform computed for our SLy configuration using 
the \texttt{TEOBResumS}~\cite{Nagar:2018zoe} model. 
The tidal phase estimate of the \texttt{TEOBResumS} model is slightly less attractive 
for frequencies around $1000 \ \rm Hz$, but more attractive at higher frequencies. 
Finally, the original \texttt{NRTidal} model is shown as a green line.
The tidal contribution is overestimated at about $f\sim1000 \ \rm Hz$, and later underestimated.
This oscillatory behavior has been seen before,
e.g., \cite{Dietrich:2017aum,Dietrich:2018upm,Dietrich:2018uni},
and could potentially lead to biases in the estimate of
tidal effects from GW signals~\cite{Dudi:2018jzn}.
For both \texttt{NRTidal} and \NRTidalp the growth of the tidal phase
around merger is much smaller than for any other approximant,
which generally reduces possible pathologies in more extreme regions of the
parameters, e.g., a cancellation of the point-particle and attractive 
tidal phase close to merger.
As expected the \NRTidalp model stays within the gray shaded region and, thus, 
close to the numerical relativity dataset used for the calibration.

\subsubsection{Frequency domain phase}

\begin{figure}[t]
\includegraphics[width=0.5\textwidth]{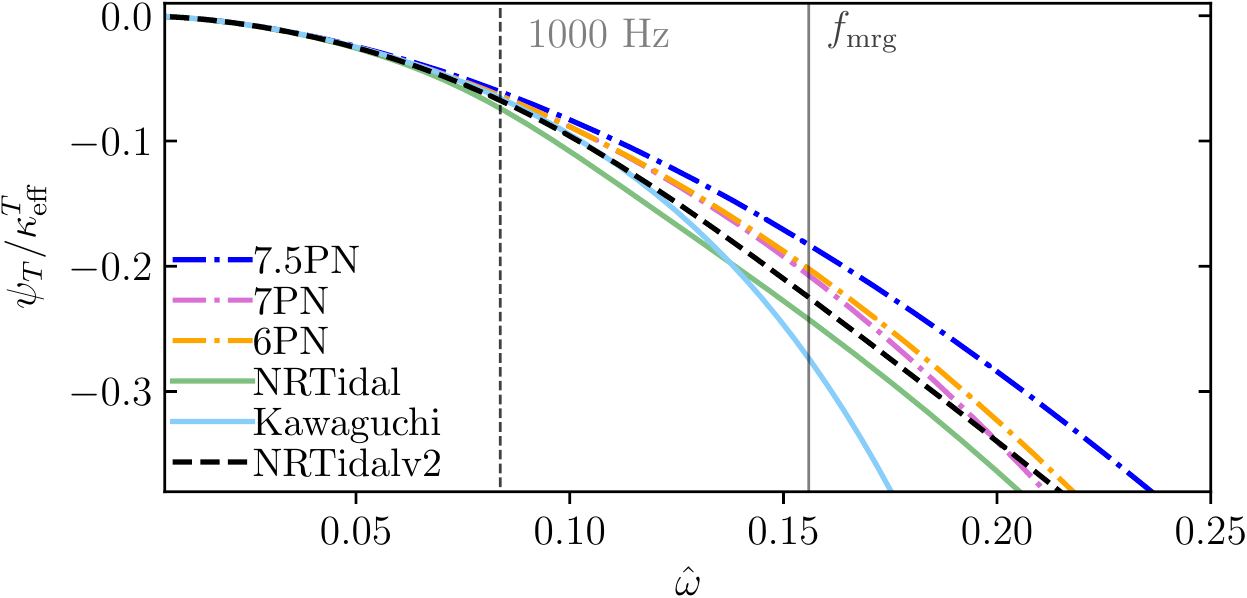}
\caption{Tidal phase in the frequency domain.
We show the 6PN (orange), 7PN (orchid), and 7.5PN (blue) tidal phase estimates,
as well as the original \texttt{NRTidal}~\cite{Dietrich:2017aum} (green)
and the Kawaguchi~\emph{et al.}~\cite{Kawaguchi:2018gvj} (cyan) approximant.
The new  model is shown with a black dashed line.
We mark the frequencies of $1000\ \rm Hz$ and the merger frequency corresponding 
to our SLy setup, described in Table~\ref{tab:calibrating_hybrids}.}
\label{fig:fit_fd}
\end{figure}

As in Ref.~\cite{Dietrich:2017aum} we will employ the stationary phase approximation (SPA), discussed in, e.g.,~\cite{Damour:2012yf},
to derive the tidal phase contribution $\psi_T$ in the frequency domain, i.e.,
we solve
\begin{equation}
  \frac{\text{d}^2 \psi_T(\omega)}{\text{d} \omega^2} =
  \frac{1}{\omega} \frac{\text{d} \phi_T(\omega)}{\text{d} \omega} \label{eq:SPA}
\end{equation}
to obtain $\psi_T$.
Although $\phi_T$ is given explicitly, we solve Eq.~\eqref{eq:SPA}
numerically and approximate the result with a Pad{\'e} approximant similar to
Eqs.~\eqref{eq:NRTidal_TD} and \eqref{eq:Pade_TD}:
\begin{equation}
\psi_T (x) = - \kappa_{\rm eff}^T
\frac{39}{ 16 \nu} x^{5/2} \tilde{P}_{\rm NRTidalv2}(x) \ ,  \label{eq:NRTidal_FD}
\end{equation}
with
\begin{small}
\begin{equation}
\tilde{P}_{\rm NRTidalv2}(x) = \frac{1 + \tilde{n}_1 x+  \tilde{n}_{3/2} x^{3/2} + \tilde{n}_2 x^2+ \tilde{n}_{5/2} x^{5/2} + \tilde{n}_3 x^3}
{1+ \tilde{d}_1 x+ \tilde{d}_{3/2} x^{3/2}+ \tilde{d}_2 x^2} \label{eq:Pade_FD},
\end{equation}
\end{small}
and
\begin{small}
\begin{subequations}\label{eq:Pade_FD_constraints}
\begin{eqnarray}
 \tilde{n}_1     & = & \tilde{c}_1 + \tilde{d}_1, \label{eq:Pade_FD_constraint1} \\
 \tilde{n}_{3/2}     & = & \frac{\tilde{c}_1 \tilde{c}_{3/2}-\tilde{c}_{5/2}-
 \tilde{c}_{3/2} \tilde{d}_1 + \tilde{n}_{5/2}}{\tilde{c}_1},  \label{eq:Pade_FD_constraint2}  \\
 \tilde{n}_{2} & = & \tilde{c}_2 + \tilde{c}_1 \tilde{d}_1 + \tilde{d}_2 ,  \label{eq:Pade_FD_constraint3}  \\
 \tilde{d}_{3/2} & = & - \frac{\tilde{c}_{5/2}+\tilde{c}_{3/2} \tilde{d}_1 - \tilde{n}_{5/2}}{\tilde{c}_1},  \label{eq:Pade_FD_constraint4} 
 \end{eqnarray}
 \end{subequations}
\end{small}
where the known coefficients are:
\begin{small}
\begin{subequations}\label{eq:Pade_FD_coeffs}
\begin{align}
  \tilde{c}_{1}   & =  \frac{3115}{1248}, \qquad
  & \tilde{c}_{3/2} & = -\pi , \label{eq:Pade_FD_coeff1} \\
  \tilde{c}_2     & =  \frac{28024205}{3302208},  \qquad
  & \tilde{c}_{5/2}  & =  - \frac{4283 \pi}{1092}. \label{eq:Pade_FD_coeff2}
\end{align}
\end{subequations}
\end{small}
and the fitting coefficients are:
 \begin{small}
 \begin{subequations}
\begin{align}
  \tilde{n}_{5/2} & =  90.550822, \qquad   
  & \tilde{n}_3     & = -60.253578, \\        
  \tilde{d}_1     & = -15.111208,  \qquad  
  & \tilde{d}_2     & =  8.0641096. 
      \end{align}
      \end{subequations}
\end{small}

\begin{figure}[t]
\includegraphics[width=0.5\textwidth]{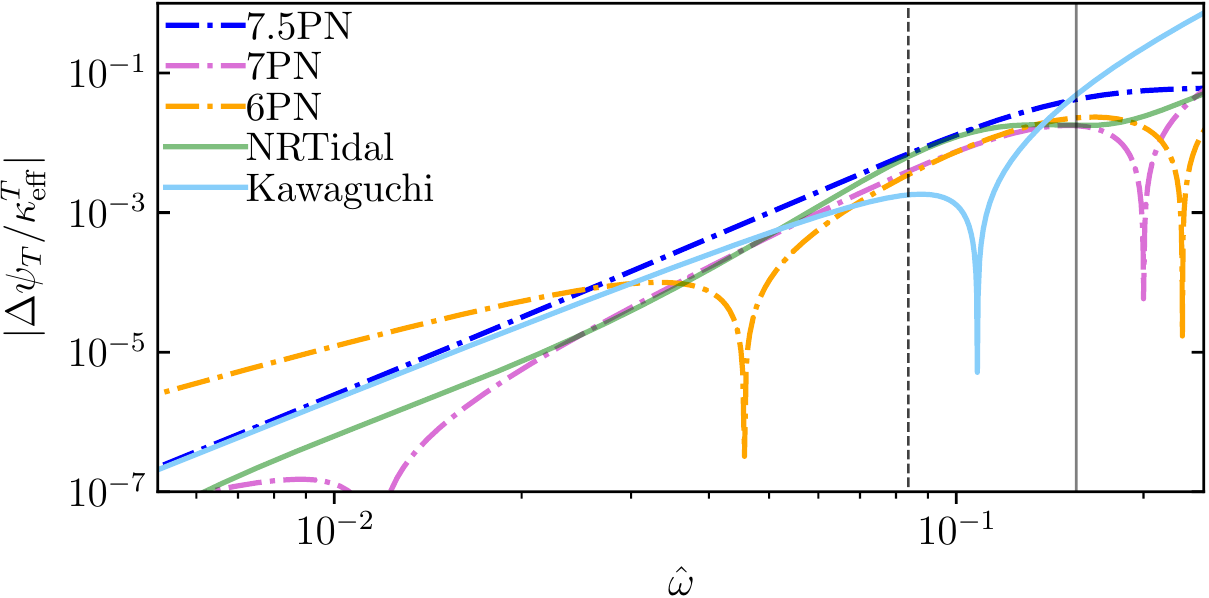}
\caption{Absolute magnitude of the tidal phase difference between
frequency domain approximants and the \NRTidalp model. 
The vertical dashed line represents $1000\ \rm Hz$, the 
frequency up to which the Kawaguchi~\emph{et al.}\ model was calibrated, 
and the solid line marks the merger frequency of the SLy setup we consider.}
\label{fig:fit_fd_diff}
\end{figure}

We present the final tidal phase contribution in
the frequency domain in Fig.~\ref{fig:fit_fd} for a number of
different GW approximants.
Fig.~\ref{fig:fit_fd_diff} shows the corresponding phase differences
with respect to the \NRTidalp model on a double logarithmic scale.

\subsection{Tidal amplitude corrections}
\label{sec:tidal_ampl}
\label{sec:improvements:tidal_ampl}

\begin{figure}[t]
\includegraphics[width=0.5\textwidth]{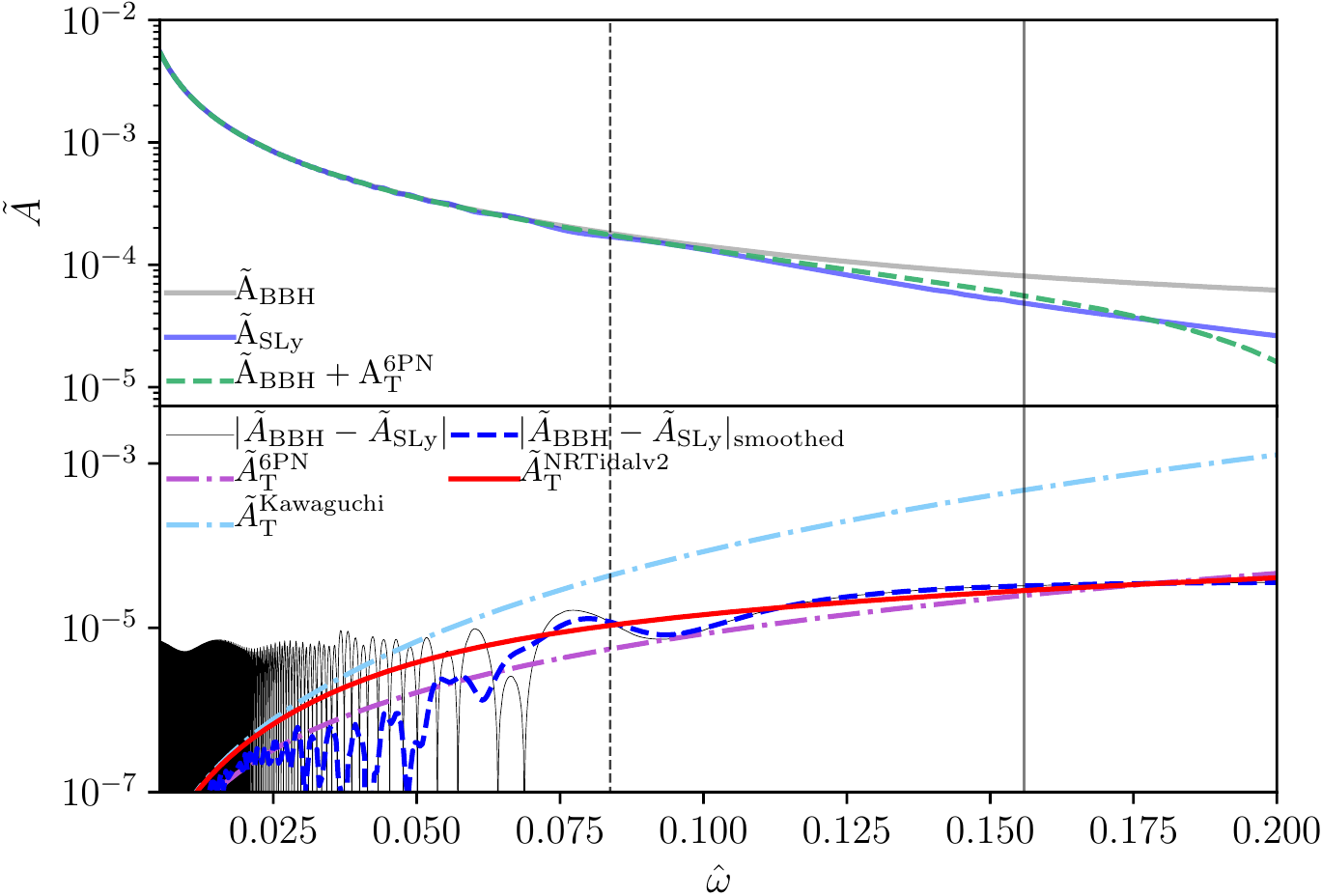}
\caption{Frequency domain waveform amplitude.
Top panel: We show the BBH and BNS hybrids' amplitudes as well as the BBH amplitude augmented with the
6PN tidal terms presented in Eq.~\eqref{eq:A_FD_6PN}.
Bottom panel: amplitude differences with the raw data in black and the smoothed
data as a blue dashed line. 
The final \NRTidalp fit is shown in red. The vertical dashed lines refer to $1000\ \rm Hz$
and to the merger frequency of the BNS hybrid respectively.}
\label{fig:fit_fd_amp}
\end{figure}

The extraction of binary properties relies mostly
on the GW phase, which makes an accurate description of $\psi$
the primary target of GW modeling.
However, a realistic estimate of the GW amplitude is also 
of importance, e.g., for a precise distance measurement.

Therefore, we will discuss a possible extension of the \NRTidal
approach including a tidal amplitude correction in the frequency domain.
An alternative time and frequency domain amplitude correction is presented
in Appendix~\ref{app:tidal_amp}.

Here, we will derive the frequency domain tidal correction from the frequency domain
representation of the SLy and BBH \texttt{TEOBResumS}-NR hybrids, described in Table~\ref{tab:calibrating_hybrids}.
We do not employ the H4 and MS1b setup for the amplitude correction 
since their lower merger frequencies add additional 
complications during the construction procedure.
The top panel of Fig.~\ref{fig:fit_fd_amp} shows $\tilde{A}$ for our generic setups and also the
BBH result augmented ($\tilde{A}=\tilde{A}_{\rm BBH} + \tilde{A}_{\rm T}$) by the 6PN expression:
\begin{equation}
 \tilde{A}_T^{\rm 6PN} =
 \sqrt{\frac{160 \pi \nu}{27}} \frac{M^2}{D_L} \
 {\kappa^T_{\rm eff}} x^{-7/4} \left(-\frac{27}{16}x^5-\frac{449}{64}x^6\right),
 \label{eq:A_FD_6PN}
\end{equation}
e.g., Ref.~\cite{Hotokezaka:2016bzh}, where $D_L$ is the luminosity distance of the source, which is the appropriate substitution
for the effective distance used in Ref.~\cite{Hotokezaka:2016bzh} for our case.%
\footnote{Note further that as before, we have restricted our analysis to the
leading order mass ratio effect and do not incorporate further mass ratio dependence in the PN parameters.
Furthermore, we restrict our consideration to gravitoelectric
contributions and do not consider gravitomagnetic
tidal effects recently computed in~\cite{Banihashemi:2018xfb}.}
Kawaguchi~\emph{et al.}~\cite{Kawaguchi:2018gvj} extended Eq.~\eqref{eq:A_FD_6PN} to
\begin{small}
\begin{equation}
\begin{split}
 \tilde{A}_T^{\rm Kawaguchi} &=
 \sqrt{\frac{160 \pi \nu}{27}} \frac{M^2}{D_L} \
 {\kappa^T_{\rm eff}} x^{-7/4} \\
&\quad \times  	\left(-\frac{27}{16}x^5-\frac{449}{64}x^6-4251 x^{7.890}\right).
 \label{eq:A_FD_Kawaguchi}
\end{split}
\end{equation}
\end{small}
Based on the good agreement we have found between the results of Ref.~\cite{Kawaguchi:2018gvj}
and the new \NRTidalp phase description below $1000\ \rm Hz$, we want to use Eq.~\eqref{eq:A_FD_Kawaguchi}
as baseline for a possible frequency amplitude extension of the \NRTidalp
approximant.\footnote{We note that the phase and amplitude extension presented in~\cite{Kawaguchi:2018gvj}
follow different approaches: While the tidal phase correction is based on an additional contribution due to
non-linear tides, i.e., a higher order $\tilde{\Lambda}$ contribution, the amplitude correction
only uses linear tidal effects, but adds an effectively higher order PN coefficient.
Therefore, the proposed amplitude extension of~\cite{Kawaguchi:2018gvj} can easily be incorporated
in our approach.}

For this purpose we employ the ansatz
\begin{small}
\begin{equation}
 \tilde{A}_T^{\rm NRTidalv2} =
 - \sqrt{\frac{5 \pi \nu}{24}} \frac{9 M^2 }{D_L}
 \kappa^T_{\rm eff} x^{13/4} \frac{1 + \frac{449}{108} x + \frac{22672}{9} x^{2.89}}{1+d \ x^4}.
 \label{eq:A_FD_NRTP}
\end{equation}
\end{small}
Equation~\eqref{eq:A_FD_NRTP} ensures that for small frequencies Eq.~\eqref{eq:A_FD_Kawaguchi}
is recovered, but that the high frequency behavior ($f>1000\ {\rm Hz}$) can be adjusted.
We obtain $d=13477.8$ by fitting the data presented in Fig.~\ref{fig:fit_fd_amp} (blue dashed line in
the bottom panel). \\

As for the previous \texttt{NRTidal} implementation, 
we add a Planck taper~\cite{McKechan:2010kp}
to end the inspiral waveform.
The taper begins at the estimated merger frequency [Eq.~(11) of~\cite{Dietrich:2018uni}]
and ends at $1.2$ times the merger frequency. 
Thus, the final amplitude is given as: 
\begin{equation}
 \tilde{A} = (\tilde{A}_{\rm BBH} + \tilde{A}_T^{\rm NRTidalv2}) \times \tilde{A}_{\rm Planck}.
\end{equation}
Because of the smooth frequency and amplitude evolution even after the moment of merger,
this taper only introduces negligible errors and does not lead to biases in the parameter estimation
of even SNR 100 signals, as shown using an injection of an SLy hybrid with the same parameters
as those considered here in~\cite{Dudi:2018jzn}. 

\subsection{Incorporating higher-order spin-spin effects}
\label{sec:improvements:spin}

\begin{figure}[t]
\includegraphics[width=0.5\textwidth]{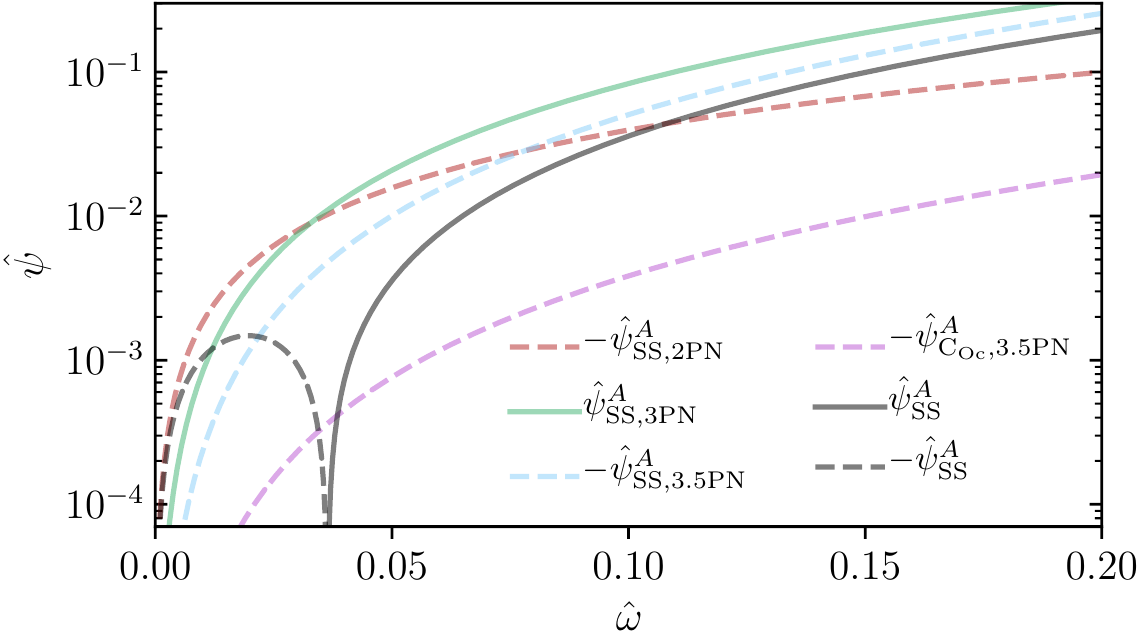}
\caption{Quadrupolar and octupolar spin-spin and cubic-in spin contributions
at 2PN, 3PN, and 3.5PN order, as well as their sum, all without the overall Newtonian scaling.
Negative terms are shown with dashed lines, positive terms with solid lines.
The plot assumes an equal-mass system with dimensionless spins $\chi_A=\chi_B=0.2$ and tidal deformabilities
of $\Lambda_A=\Lambda_B=350$. The quadrupole and octupole moments are computed according to Eq.~\eqref{eq:quad}
and Eq.~\eqref{eq:oct} ($C_{\rm Q}^A=C_{\rm Q}^B=4.30$, $C_{\rm Oc}^A=C_{\rm Oc}^B=7.28$).}
\label{fig:spin_fd}
\end{figure}

While nonspinning NSs and black holes only have a
nonzero monopole moment, spinning neutron stars and black holes
have an infinite series of nonzero (Geroch-Hansen) multipole moments,
e.g., Refs.~\cite{Pappas:2013naa, Yagi:2014bxa}.
The contributions from the stars' (mass) monopole and (spin) dipole to the binary's motion
are explicitly accounted for in the BBH baseline.
Additionally, contributions from higher spin-induced multipoles in the BBH baseline model
are indirectly included due to the calibration to NR simulations.
However, without further adjustment, all multipoles would be
specialized to the black hole values, which (as shown in~\cite{Dietrich:2018uni})
noticeably reduces the accuracy for spinning BNS systems.
Thus, to improve the performance of \NRTidalp for spinning configurations,
we include an EOS dependence in
the quadrupole and octupole, as these are the moments that appear
in current PN calculations.
These two moments (in their scalar versions) can be written as
$M^{(A,B)}_2 = -C_{\rm Q}^{A} M_{A,B}^3\chi_{A,B}^2$,
$S_3^{(A,B)} = -C_{\rm Oc}^{A,B} M_{A,B}^4\chi_{A,B}^3$,
respectively, for star $A$ and $B$.
Here $C_{\rm Q}^{A,B}$ and $C_{\rm Oc}^{A,B}$ are the
quadrupolar and octupolar spin-induced deformabilities for the individual stars.
Both $C_{\rm Q}$ and $C_{\rm Oc}$ are $1$ for a black hole.

In this paper, we extend the existing LALSuite implementation, which currently contains the EOS dependence of the quadrupole
moment only up to 3PN~\cite{Bohe:2015ana,Mishra:2016whh} to include the $3.5$PN spin-squared terms, completed using the recently
computed 3.5PN tail terms~\cite{Nagar:2018plt}. We also include leading order spin-cubed terms entering at 3.5PN order.

The contributions of the quadrupole and octupole deformations of the stars to the binary's binding
energy and energy flux have been computed through $3.5$PN,
Refs.~\cite{Bohe:2015ana,Marsat:2014xea,Nagar:2018zoe}, building on earlier work reviewed in~\cite{Blanchet:2013haa}.
We compute the phase in the frequency domain using the SPA.
These contributions to the phase were already presented in~\cite{Krishnendu:2017shb},
except the $3.5$PN spin-spin terms,
as \cite{Krishnendu:2017shb} did not have the $3.5$PN spin-squared tail term from Ref.~\cite{Nagar:2018zoe}.
Explicitly, the self-spin (i.e., $C_{\rm Q}$ and $C_{\rm Oc}$)
terms in the phasing that we add to the BBH baseline are
\begin{equation}
\begin{split}
\psi_\text{SS} &=  \frac{3 x^{-5/2}}{128 \nu} \left(\hat{\psi}^{(A)}_\text{SS, 2PN} x^2 +
\hat{\psi}^{(A)}_\text{SS, 3PN} x^3 + \hat{\psi}^{(A)}_\text{SS, 3.5PN} x^{7/2}\right)\\
&\quad + [ A \leftrightarrow B ]
\end{split}
\end{equation}
with
\begin{equation}
\begin{split}
\hat{\psi}^{(A)}_\text{SS, 2PN} &:= -50\hat{C}_{\rm Q}^{A} X_A^2\chi _A^2\\
\hat{\psi}^{(A)}_\text{SS, 3PN} &:= \frac{5}{84} \left(9407 + 8218 X_A - 2016 X_A^2\right) \hat{C}_{\rm Q}^A X_A^2\chi _A^2\\
\hat{\psi}^{(A)}_\text{SS, 3.5PN} &:= 10\biggl[\left(X_A^2 + \frac{308}{3}X_A\right)\chi _A + \left(X_B^2 - \frac{89}{3} X_B\right) \chi _B\\
&\;\quad - 40 \pi\biggr]\hat{C}_{\rm Q}^{A} X_A^2 \chi _A^2 - 440 \hat{C}_{\rm Oc}^{A} X_A^3 \chi _A^3
\end{split}
\end{equation}
Here we use $\hat{C}_{\rm Q}^{A} := C_{\rm Q}^{A} - 1$ and
$\hat{C}_{\rm Oc}^{A} := C_{\rm Oc}^{A} - 1$
to remove the contribution from the black hole multipoles already present in the baseline BBH phase.

Finally, we relate $C_{\rm Q}^{A}$ to the tidal deformability $\Lambda_A$ and $C_{\rm Oc}^{A}$ to
$C_{\rm Q}^{A}$ using the EOS-insensitive relations (Tables~1 and 2 from~\cite{Yagi:2016bkt}):
\begin{equation}
\begin{split}
 \log(C_{\rm Q}^A) &= 0.1940 + 0.09163 \log(\Lambda_A) + 0.04812 \log^2(\Lambda_A) \\
             &\quad - 0.004286 \log^3(\Lambda_A) + 0.00012450 \log^4(\Lambda_A)
             \label{eq:quad}
\end{split}
\end{equation}
and
\begin{equation}
\begin{split}
 \log(C_{\rm Oc}^A) &= 0.003131 + 2.071 \log(C_{\rm Q}^A) -0.7152 \log^2(C_{\rm Q}^A) \\
   &\quad + 0.2458 \log^3(C_{\rm Q}^A) -0.03309 \log^4(C_{\rm Q}^A). \label{eq:oct}
\end{split}
\end{equation}

To allow a better interpretation of the spin-spin terms discussed above,
we present in Fig.~\ref{fig:spin_fd} the individual contributions
$\hat{\psi}^{(A)}_\text{SS, 2PN}$,
$\hat{\psi}^{(A)}_\text{SS, 3PN}$,
and $\hat{\psi}^{(A)}_\text{SS, 3.5PN}$.
In addition, for better visibility, we also show explicitly
the spin-cubed octupole term $\hat{\psi}^{(A)}_{C_{\rm Oc}, 3.5\text{PN}} = - 440 \hat{C}_{\rm Oc}^{A} X_A^3 \chi _A^3$.
For an equal mass setup with $\Lambda^{A,B}=350$ and $\chi^{A,B}=0.2$ the 2PN contribution dominates
up to $\hat{\omega} \sim 0.06$, before the positive 3PN term becomes larger.
Overall, we find that except the 3PN contribution all terms are negative for the chosen setup.
We also see that throughout the inspiral the octupole term is about 1 order of magnitude smaller than
other contributions. 
This observation remains valid even for spins close to break-up $\chi \sim 0.75$.
Thus, we do not attempt to include additional higher order multipoles.\\

\subsection{Precession dynamics}

We conclude the discussion of the model by shortly describing the
incorporation of precession. The precession dynamics in \texttt{IMRPhenomPv2\_NRTidalv2}
is included as in the previous \texttt{IMRPhenomPv2\_NRTidal} approach~\cite{Dietrich:2018uni}.
For this we assume that the spin-orbit coupling can be
approximately separated into components
parallel and perpendicular to the instantaneous
orbital angular momentum, where the component perpendicular to
the orbital angular momentum is driving the precessional
motion~\cite{Kidder:1995zr,Apostolatos:1994mx,Buonanno:2002fy,
Schmidt:2010it,Schmidt:2012rh,Schmidt:2014iyl}.

Consequently, we construct a precessing tidal waveform approximant
from the spin-aligned model after adding all tidal corrections to
the underlying spin-aligned point particle model.
We then rotate the waveform to account for precession, as discussed in Refs.~\cite{Schmidt:2012rh,Schmidt:2014iyl}.

\section{Validation}
\label{sec:validation}

\begin{table*}
  \centering
  \caption{NR BNS configurations for validation of the time domain phasing.
    The columns refer to the \texttt{CoRe}-ID of the setup,
    the EOS (see Ref.~\cite{Read:2008iy}),
    the NSs' individual masses $M_{A,B}$,
    the stars' dimensionless spins $\chi_{A,B}$,
    the tidal deformabilities of the stars $\Lambda_{A,B}$,
    the tidal deformability of the binary $\tilde{\Lambda}$,
    the grid resolution covering the NS, and
    the residual eccentricity of the configuration.
    In the last column we state whether we employ Richardson extrapolation
    for a better estimate of the phase. For those setups the errors shown
    in Fig.~\ref{fig:NR_comparison} present a conservative error measure 
    and are shown as green shaded regions.
    We note that setups CoRe:BAM:0037 and CoRe:BAM:0064 have 
    also been employed for the calibration of the model; 
    see Eq.~\eqref{eq:NRdata_construction}.
    }
\begin{tabular}{l||cccccccc|ccc}
\hline
\hline
  Name & EOS & $M_{A}$ $[M_\odot]$ & $M_B$ $[M_\odot]$ & $\chi_{A}$ & $\chi_{B}$ & $\Lambda_A$ & $\Lambda_B$
  & $\tilde{\Lambda}$ & $h_{\rm fine}$ $[M_\odot]$ & $e$ $[10^{-3}]$ & Richardson \\
\hline
CoRe:BAM:0001 & 2B   & 1.371733 & 1.371733 & 0.000  & 0.000  & 126.73 & 126.73 & 126.73 & 0.0930 & 7.1 & \xmark \\
CoRe:BAM:0011 & ALF2 & 1.500006 & 1.500006 & 0.000  & 0.000  & 382.77 & 382.77 & 382.77 & 0.1250 & 3.1 & \xmark \\
CoRe:BAM:0037 & H4   & 1.371733 & 1.371733 & 0.000  & 0.000  & 1006.2 & 1006.2 & 1006.2 & 0.0833 & 0.9 & \cmark \\
CoRe:BAM:0039 & H4   & 1.372588 & 1.372588 & 0.141  & 0.141  & 1001.8 & 1001.8 & 1001.8 & 0.0833 & 0.5 & \cmark \\
CoRe:BAM:0062 & MS1b & 1.350398 & 1.350398 & -0.099 & -0.099 & 1531.5 & 1531.5 & 1531.5 & 0.0970 & 1.8 & \cmark \\
CoRe:BAM:0064 & MS1b & 1.350032 & 1.350032 & 0.000  & 0.000  & 1531.5 & 1531.5 & 1531.5 & 0.0970 & 1.8 & \cmark \\
CoRe:BAM:0068 & MS1b & 1.350868 & 1.350868 & 0.149  & 0.149  & 1525.2 & 1525.2 & 1525.2 & 0.0970 & 2.3 & \cmark \\
CoRe:BAM:0081 & MS1b & 1.500016 & 1.000001 & 0.000  & 0.000  & 863.8  & 7022.3 & 2425.5 & 0.1250 & 15. & \xmark \\
CoRe:BAM:0094 & MS1b & 1.944006 & 0.944024 & 0.000  & 0.000  & 182.9  & 9279.9 & 1308.2 & 0.1250 & 3.4 & \xmark \\
CoRe:BAM:0105 & SLy  & 1.350608 & 1.350608 & 0.106  & 0.106  & 388.2  & 388.24 &  388.2 & 0.0783 & 0.7 & \cmark \\
\hline
\end{tabular}
 \label{tab:NRconfigs}
\end{table*}

\begin{figure*}[t]
\includegraphics[width=0.495\textwidth]{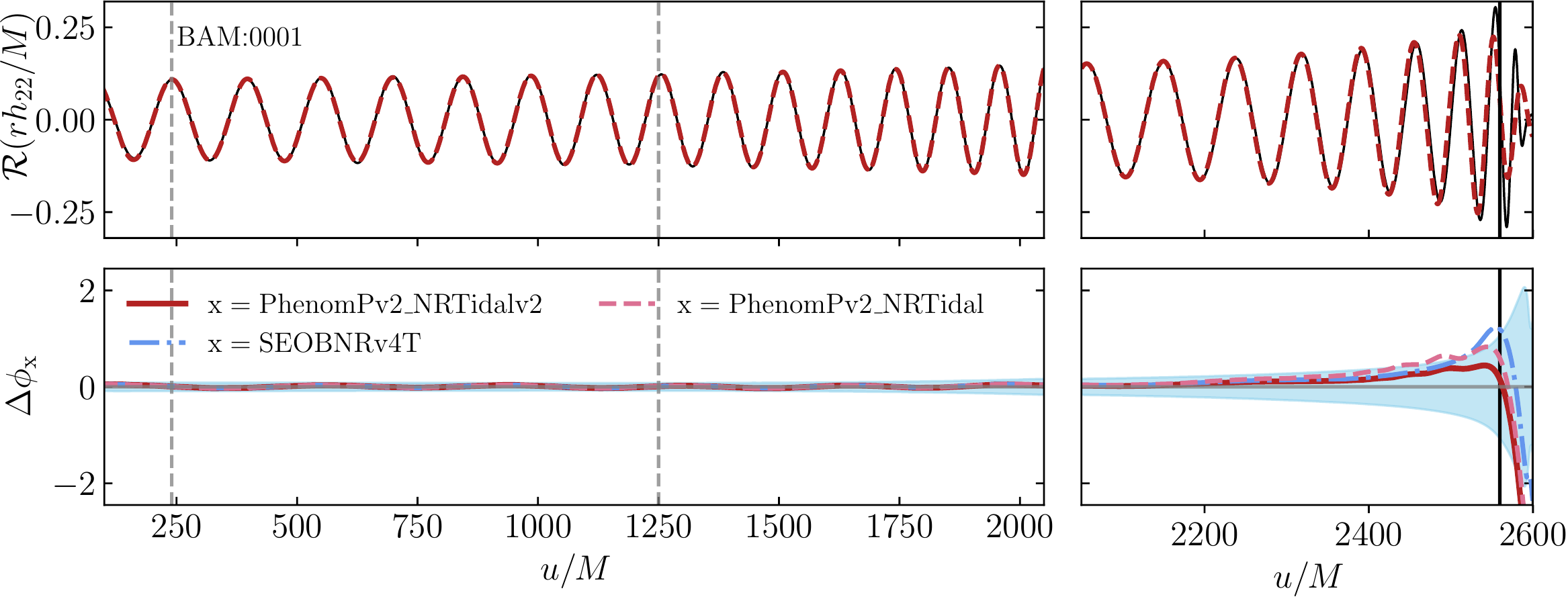}
\includegraphics[width=0.495\textwidth]{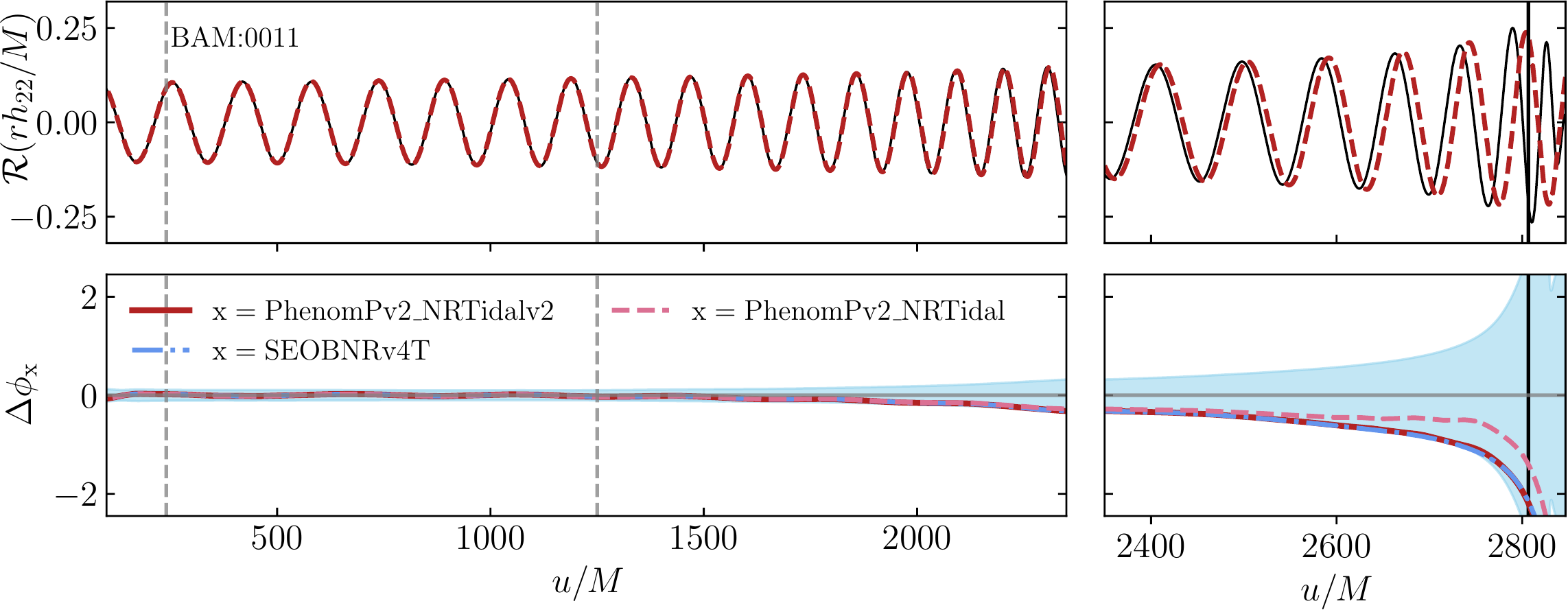}
\includegraphics[width=0.495\textwidth]{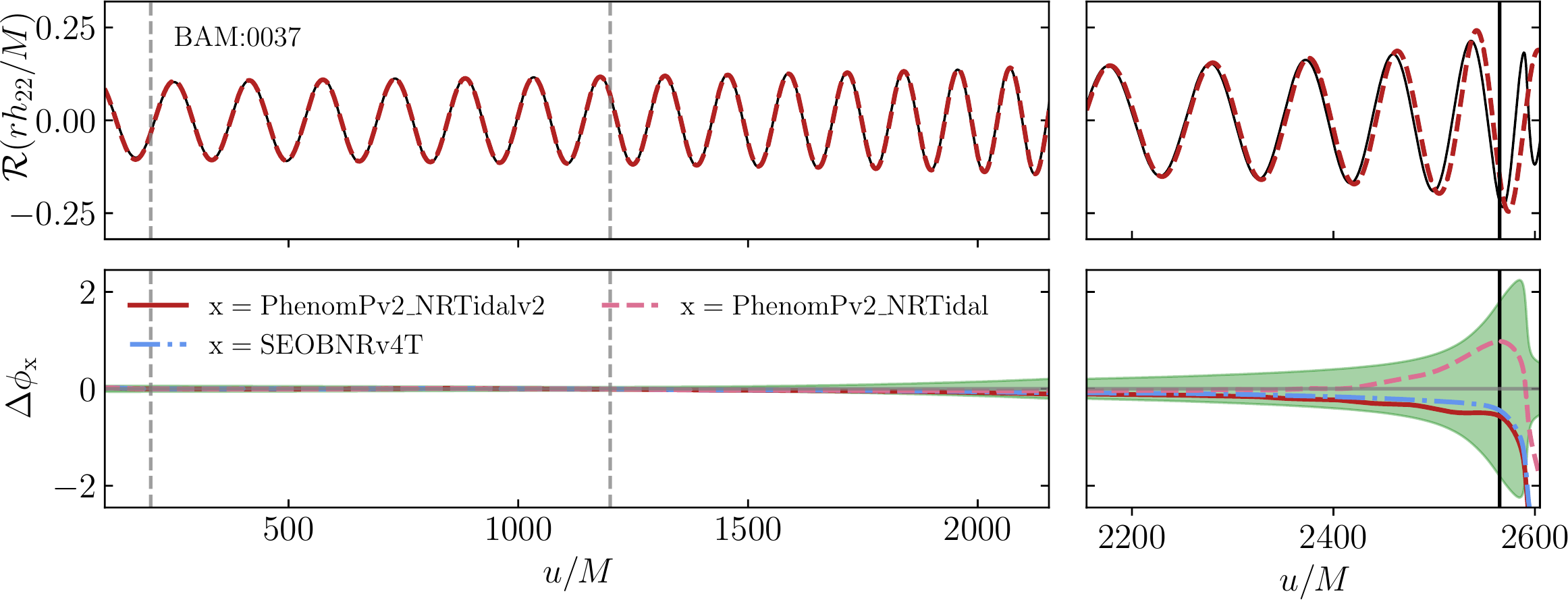}
\includegraphics[width=0.495\textwidth]{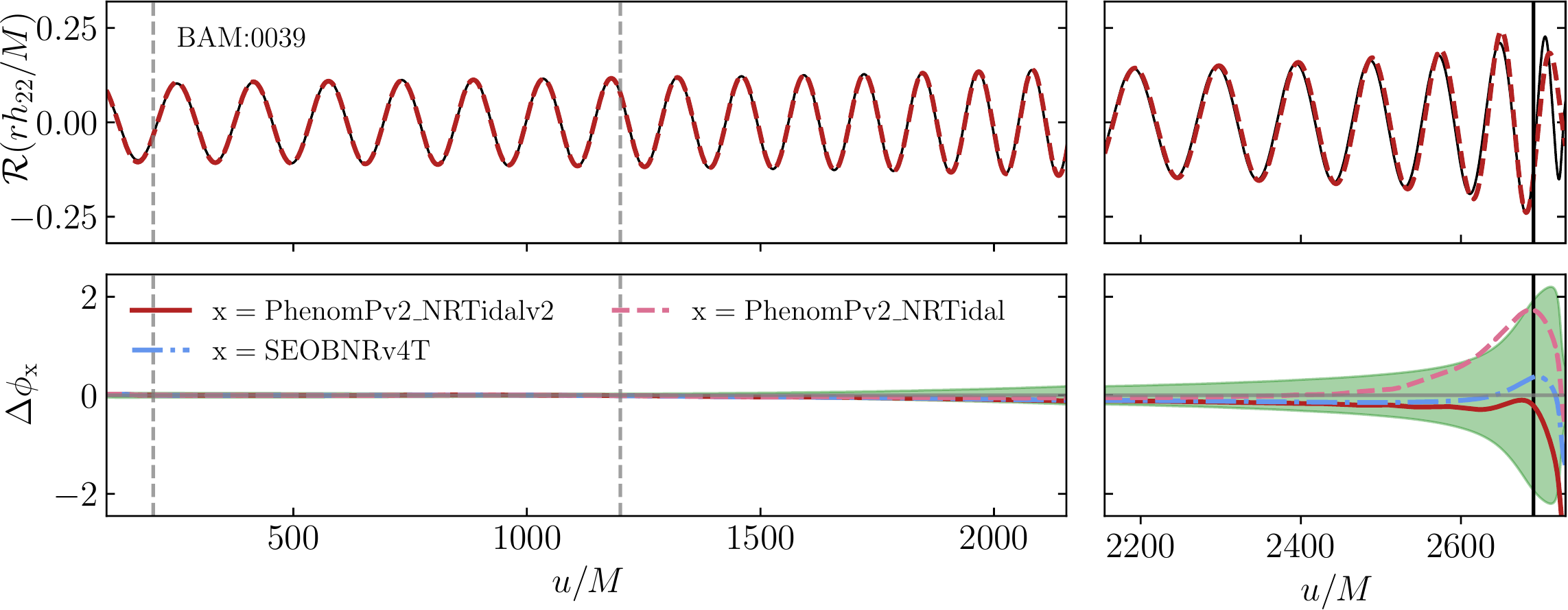}
\includegraphics[width=0.495\textwidth]{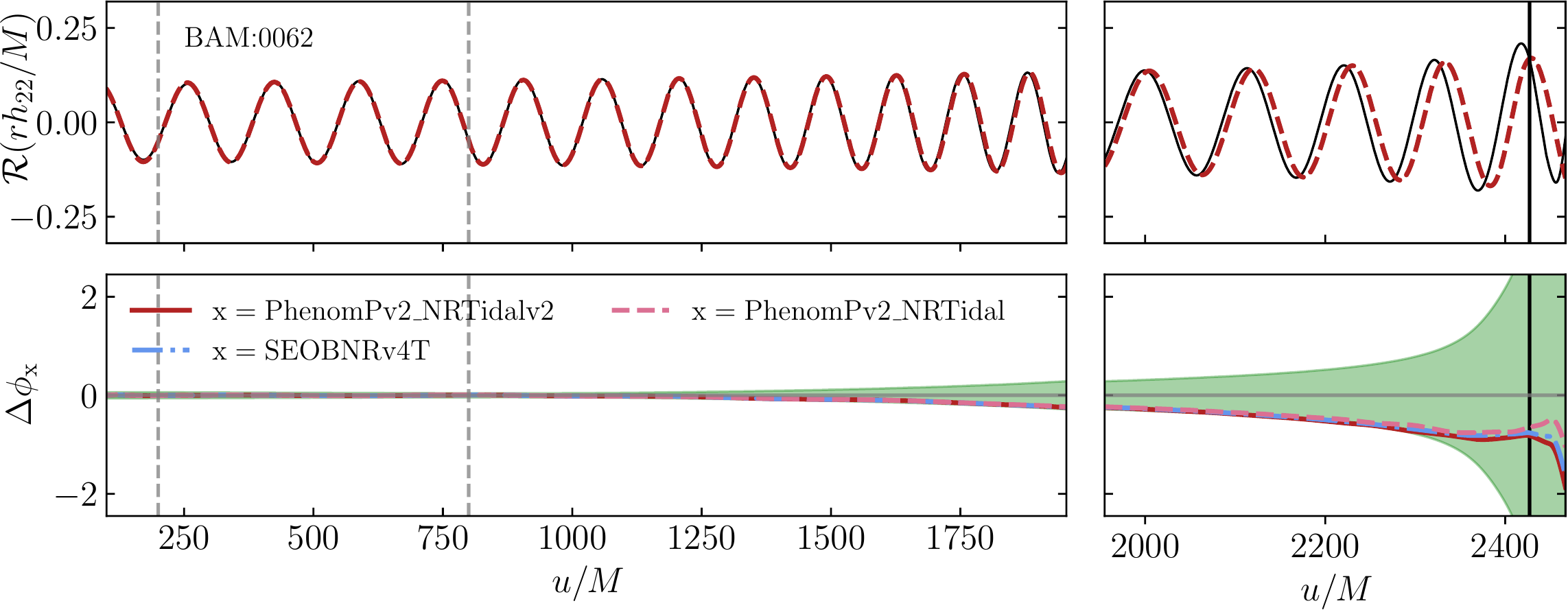}
\includegraphics[width=0.495\textwidth]{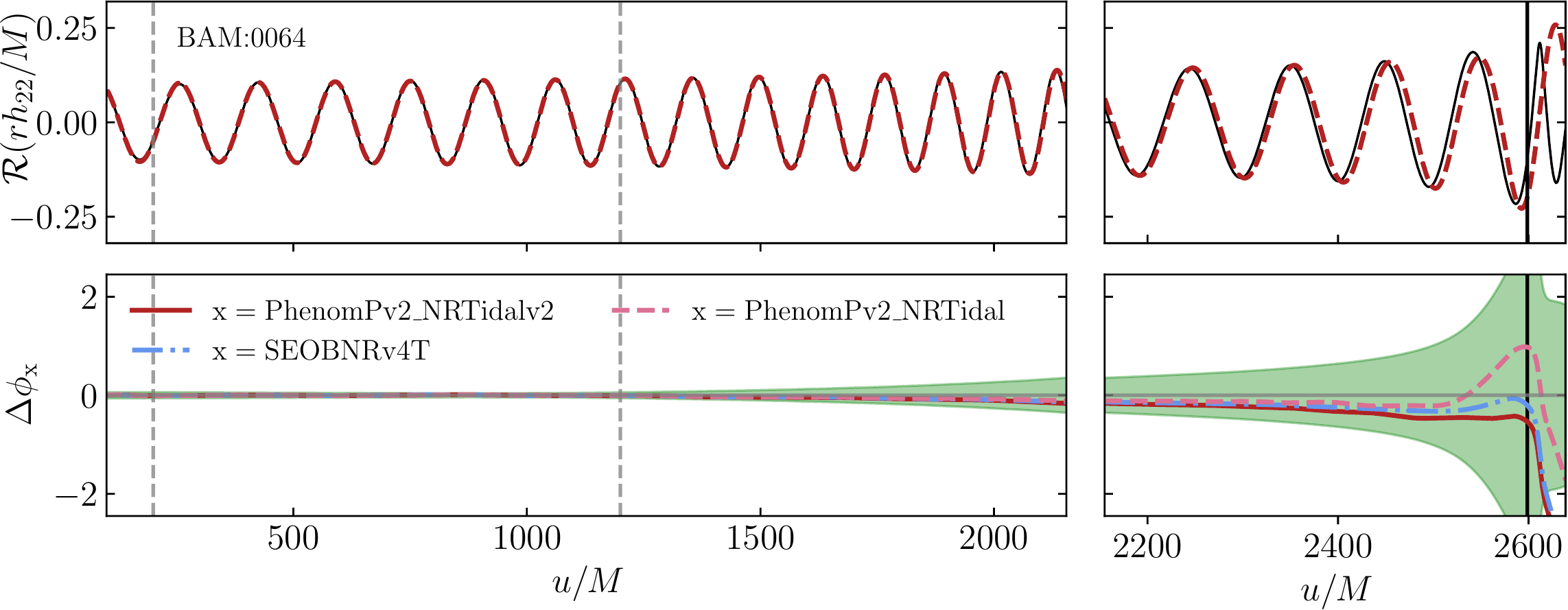}
\includegraphics[width=0.495\textwidth]{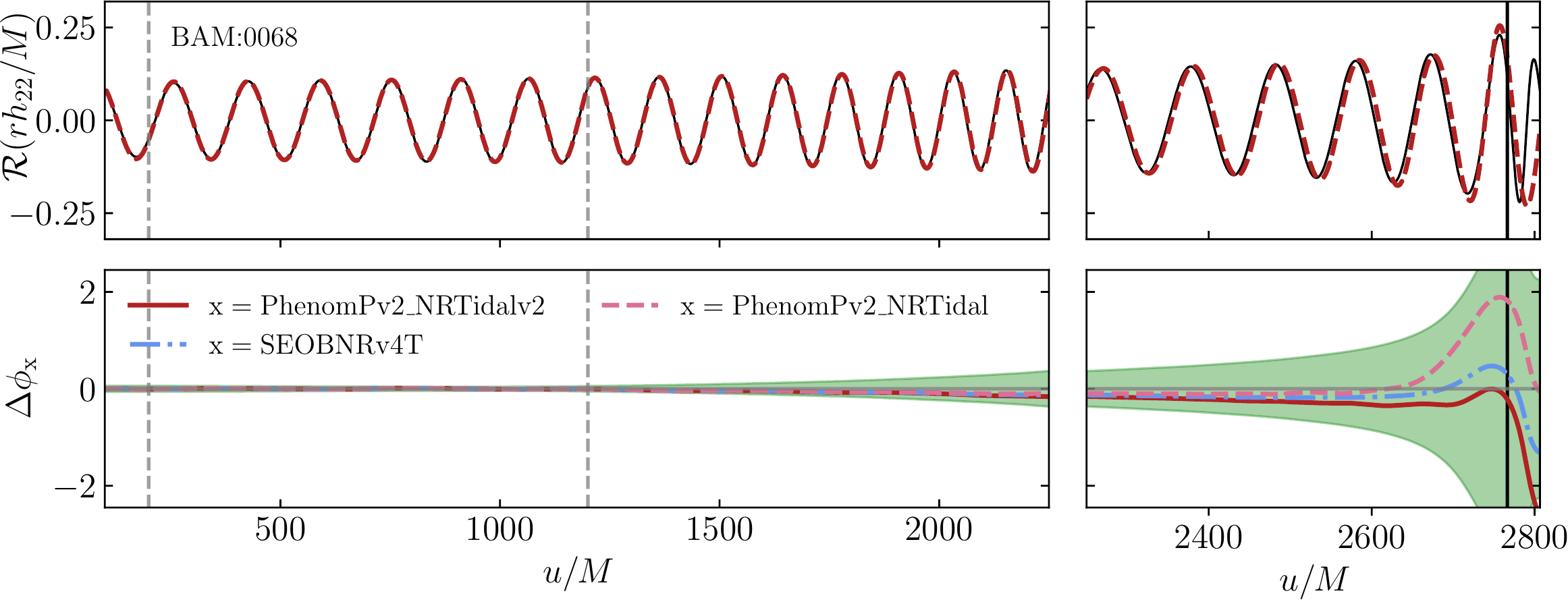}
\includegraphics[width=0.495\textwidth]{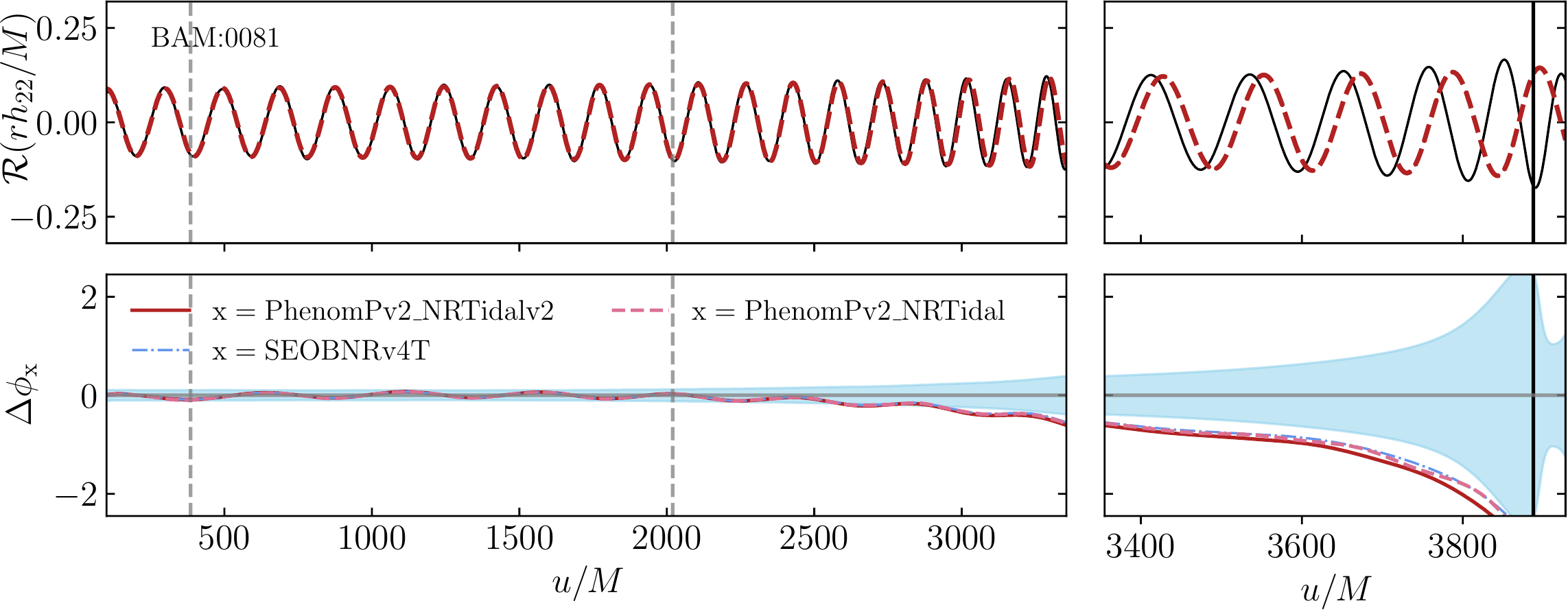}
\includegraphics[width=0.495\textwidth]{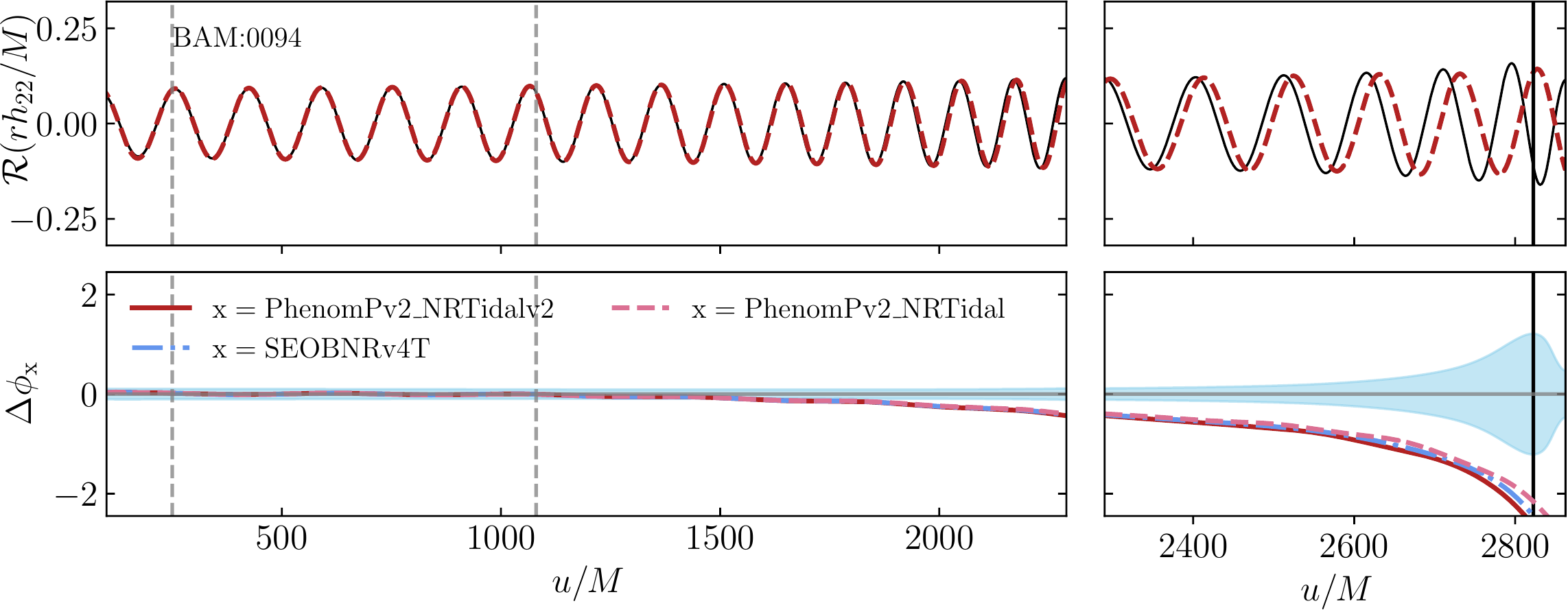}
\includegraphics[width=0.495\textwidth]{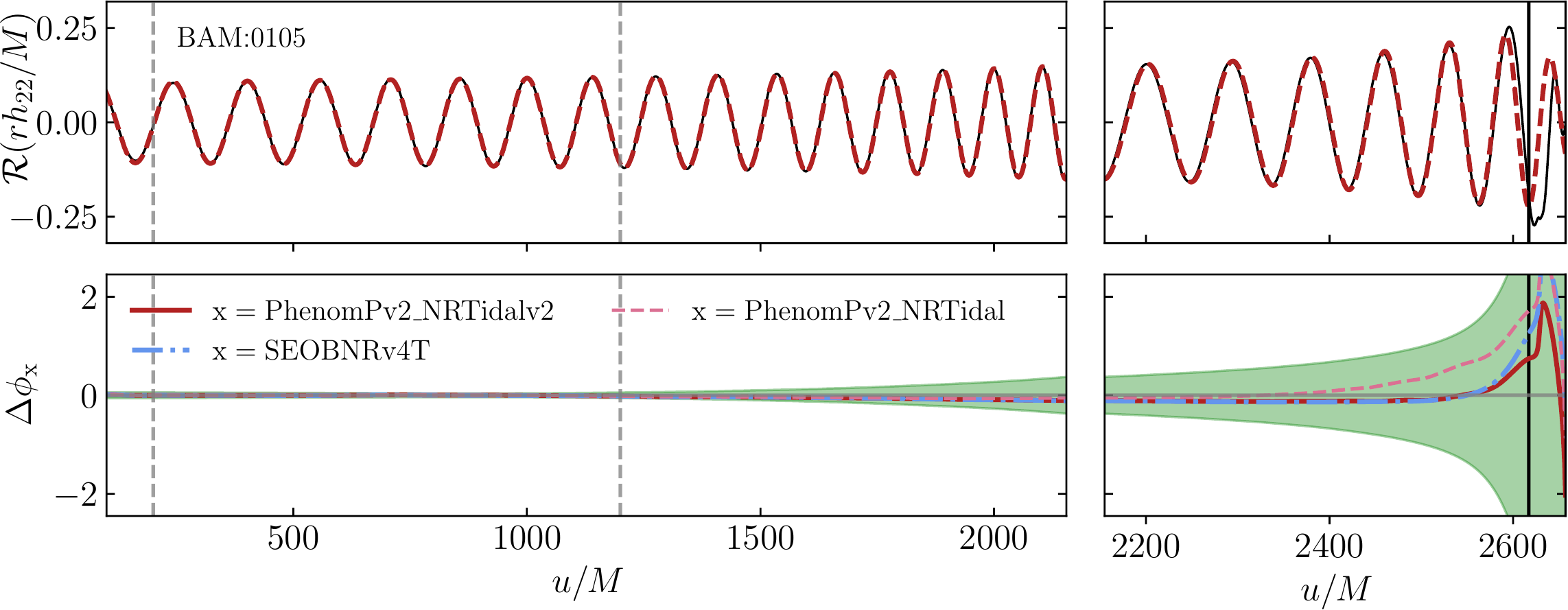}
\caption{Top panel: Real part of the GW signal obtained from the NR data (black) and the 
\texttt{IMRPhenomPv2\_NRTidalv2} model (red dashed). 
Bottom panel: Phase difference between the NR data listed in Table~\ref{tab:NRconfigs} 
and the \texttt{IMRPhenomPv2\_NRTidalv2},
\texttt{IMRPhenomPv2\_NRTidal}, and \texttt{SEOBNRv4T} models; note that we discard the \texttt{IMR}-prefix 
for better visibility in the panel legends. 
The green shaded regions denote the errors computed using Richardson extrapolation and $u$ is the retarded time, 
while blue shaded regions present the phase difference between the two highest NR resolutions. 
The vertical dashed lines mark the alignment region while the solid vertical line marks the merger. 
See the main text for further details.}
\label{fig:NR_comparison}
\end{figure*}

\subsection{Time domain comparison with NR simulations}
\label{sec:validation:NR}

As a first validation check, we compute the time domain phase difference
between \texttt{IMRPhenomPv2\_NRTidalv2} and a selected set of NR data; 
see Table~\ref{tab:NRconfigs}.
All of the employed waveforms are publicly available in the \texttt{CoRe} database
(\texttt{www.computational-relativity.org}~\cite{Dietrich:2018phi}).
In addition to \texttt{IMRPhenomPv2\_NRTidalv2},
we also present the phase difference with respect to
\texttt{SEOBNRv4T} and \texttt{IMRPhenomPv2\_NRTidal} 
in Fig.~\ref{fig:NR_comparison}.

\textbf{Waveform alignment:}
For comparison, we align all waveforms with respect to the NR
data by minimizing the phase difference in the time interval $[t_i,t_f]$
\begin{equation}
\mathcal{I}(\delta t, \delta \phi) = \int_{t_i}^{t_f}|\phi_{\rm NR}(t) -
\phi_{\rm x}(t+\delta t) + \delta \phi| \, \text{d} t \label{eq:alignment},
\end{equation}
where $\rm x$ denotes the individual waveform approximant.
The alignment windows are marked by vertical dashed 
lines in Fig.~\ref{fig:NR_comparison}.

\textbf{NR data uncertainty:}
For a quantitative comparison with respect to the NR data,
we assign each dataset with an uncertainty, where we generally
distinguish between (i) setups employing the high-order flux scheme of~\cite{Bernuzzi:2016pie}
for which clean convergence is found throughout the inspiral,
and (ii) setups whose behavior is monotonic, but no clean convergence is present.
For the setups employing the high-order flux scheme,
we obtain a better phase estimate and an error measure (green shaded region) due
to Richardson extrapolation~\cite{Bernuzzi:2016pie,Dietrich:2018upm}; cf.~Sec.~\ref{sec:NR}.
Other configurations are marked by blue shaded regions. 
For these cases, 
the uncertainty due to numerical discretization is estimated by the difference between
the two highest resolutions, which is not necessarily a conservative error estimate. 
For both scenarios, we also include an error measuring the effect of the finite radius 
extraction of the GW from the numerical domain.
This error measure is obtained by computing the difference
in the waveform's phase with respect to different extraction radii;
see e.g.~Refs.~\cite{Bernuzzi:2016pie,Dietrich:2017feu} 
for a more detailed discussion.

\textbf{\NRtidalp dephasing:}
Considering the performance of the \NRtidalp approximation,
we find that for all cases with reliable error measure (green shaded regions),
the dephasing between the model and the
NR data is well within the error estimate and never exceeds $1\ \rm rad$.
The performance is comparable with the \texttt{SEOBNRv4T} model
which is shown as a blue dashed-dotted line.\footnote{We note that very recently 
Ref.~\cite{Lackey:2018zvw} has constructed a reduced order model of 
\texttt{SEOBNRv4T} which can also be used directly 
for parameter estimation.}
Considering the difference with respect to \texttt{IMRPhenomPv2\_NRTidal},
we find that as expected the new \NRtidalp model is less attractive, 
which is caused by the slightly different behavior in the frequency range 
$\mo \geq 0.05$.

For the NR setups which show no clear convergence throughout the inspiral,
we find that for most cases the estimated uncertainty is larger than the
phase difference between the \NRTidalp and the NR data, the exceptions
are BAM:0081 and BAM:0094.
These setups are characterized by high mass ratios
[BAM:0094 is to date the NR dataset with the largest simulated mass ratio ($q=2.1$)]
and tidal deformabilities which are in tension with the observation
of GW170817~\cite{LIGOScientific:2018mvr}.

Additional simulations with clean convergence for large mass ratios are needed to allow
an overall improvement of BNS models in these regions of the parameter space (see Appendix~\ref{app:mass_ratio}).

\subsection{Mismatch Computations with respect to EOB-NR hybrids}
\label{sec:validation:hybrids}

\begin{table*}[t]
  \centering
  \caption{\ac{BNS} hybrid configurations.
    The columns describe:
    the name of the hybrid (CoRe database ID),
    the EOS, cf.~\cite{Read:2008iy},
    the NSs' individual masses $M_{A,B}$,
    the stars' dimensionless spins $\chi_{A,B}$,
    the stars' compactnesses $C_{A,B}$,
    the tidal deformabilities of the stars $\Lambda_{A,B}$,
    the tidal deformability of the binary $\tilde{\Lambda}$,
    the effective dimensionless coupling constant
    $\kappa^T_{\rm eff}$,
    and the merger frequency $f_{\rm mrg}$.
    }
\begin{tabular}{l||cccccccccccc}
\hline
\hline
  Name & EOS & $M_{A}$ $[M_\odot]$ & $M_B$ $[M_\odot]$ & $\chi_{A}$ & $\chi_{B}$ & $C_A$& $C_B$ & $\Lambda_A$ & $\Lambda_B$ & $\tilde{\Lambda}$ & $\kappa^T_{\rm eff}$ & $f_{\rm mrg}$ [Hz] \\
     \hline
\multicolumn{12} {l} {\bfseries equal mass, non-spinning}     \\
\hline
CoRe:Hyb:0001  & 2B   & 1.3500 & 1.3500 & 0.000 & 0.000 & 0.205 & 0.205 & 127.5 & 127.5 & 127.5 & 23.9  & 2567 \\
CoRe:Hyb:0002  & SLy  & 1.3500 & 1.3500 & 0.000 & 0.000 & 0.174 & 0.174 & 392.1 & 392.1 & 392.1 & 73.5  & 2010 \\
CoRe:Hyb:0003  & H4   & 1.3717 & 1.3717 & 0.000 & 0.000 & 0.149 & 0.149 & 1013.4& 1013.4& 1013.4& 190.0 & 1535 \\
CoRe:Hyb:0004  & MS1b & 1.3500 & 1.3500 & 0.000 & 0.000 & 0.142 & 0.142 & 1536.7& 1536.7& 1536.7& 288.1 & 1405 \\
CoRe:Hyb:0005  & MS1b & 1.3750 & 1.3750 & 0.000 & 0.000 & 0.144 & 0.144 & 1389.4& 1389.4& 1389.4& 260.5 & 1416 \\
CoRe:Hyb:0006  & SLy  & 1.3750 & 1.3750 & 0.000 & 0.000 & 0.178 & 0.178 & 347.3 & 347.3 & 347.3 & 65.1  & 1978 \\
\multicolumn{12} {l} {\bfseries unequal mass, non-spinning}     \\
\hline
CoRe:Hyb:0007 & MS1b& 1.5000 & 1.0000 & 0.000 & 0.000 & 0.157 & 0.109 & 866.5 & 7041.6 & 2433.5 & 456.3 & 1113  \\
CoRe:Hyb:0008 & MS1b& 1.6500 & 1.1000 & 0.000 & 0.000 & 0.171 & 0.118 & 505.2 & 4405.9 & 1490.1 & 279.4 & 1170 \\
CoRe:Hyb:0009  & MS1b& 1.5278 & 1.2222 & 0.000 & 0.000 & 0.159 & 0.130 & 779.6 & 2583.2 & 1420.4 & 266.3 & 1301 \\
CoRe:Hyb:0010   & SLy & 1.5000 & 1.0000 & 0.000 & 0.000 & 0.194 & 0.129 & 192.3 & 2315.0 & 720.0  & 135.0 & 1504 \\
CoRe:Hyb:0011   & SLy & 1.5274 & 1.2222 & 0.000 & 0.000 & 0.198 & 0.157 & 167.5 & 732.2  & 365.6  & 68.6  & 1770 \\
CoRe:Hyb:0012   & SLy & 1.6500 & 1.0979 & 0.000 & 0.000 & 0.215 & 0.142 & 93.6  & 1372.3 & 408.1  & 76.5  & 1592 \\
\multicolumn{12} {l} {\bfseries equal mass, spinning}     \\
\hline
CoRe:Hyb:0013     & H4   & 1.3726 & 1.3726 & +0.141 & +0.141 & 0.149 & 0.149 & 1009.1& 1009.1&1009.1& 189.2 & 1605 \\
CoRe:Hyb:0014 & MS1b & 1.3504 & 1.3504 & -0.099 & -0.099 & 0.142 & 0.142 & 1534.5& 1534.5&1534.5& 287.7 & 1323 \\
CoRe:Hyb:0015  & MS1b & 1.3504 & 1.3504 & +0.099 & +0.099 & 0.142 & 0.142 & 1534.5& 1534.5&1534.5& 287.7 & 1442 \\
CoRe:Hyb:0016   & MS1b & 1.3509 & 1.3509 & +0.149 & +0.149 & 0.142 & 0.142 & 1531.8& 1531.8&1531.8& 287.2 & 1456 \\
CoRe:Hyb:0017    & SLy  & 1.3502 & 1.3502 & +0.052 & +0.052 & 0.174 & 0.174 & 392.0 & 392.0 & 392.0& 73.5  & 2025 \\
CoRe:Hyb:0018   & SLy  & 1.3506 & 1.3506 & +0.106 & +0.106 & 0.174 & 0.174 & 391.0 & 391.0 & 391.0& 73.5  & 2048 \\
     \end{tabular}
 \label{tab:hybrids}
\end{table*}

\begin{figure*}[t]
\includegraphics[width=1.\textwidth]{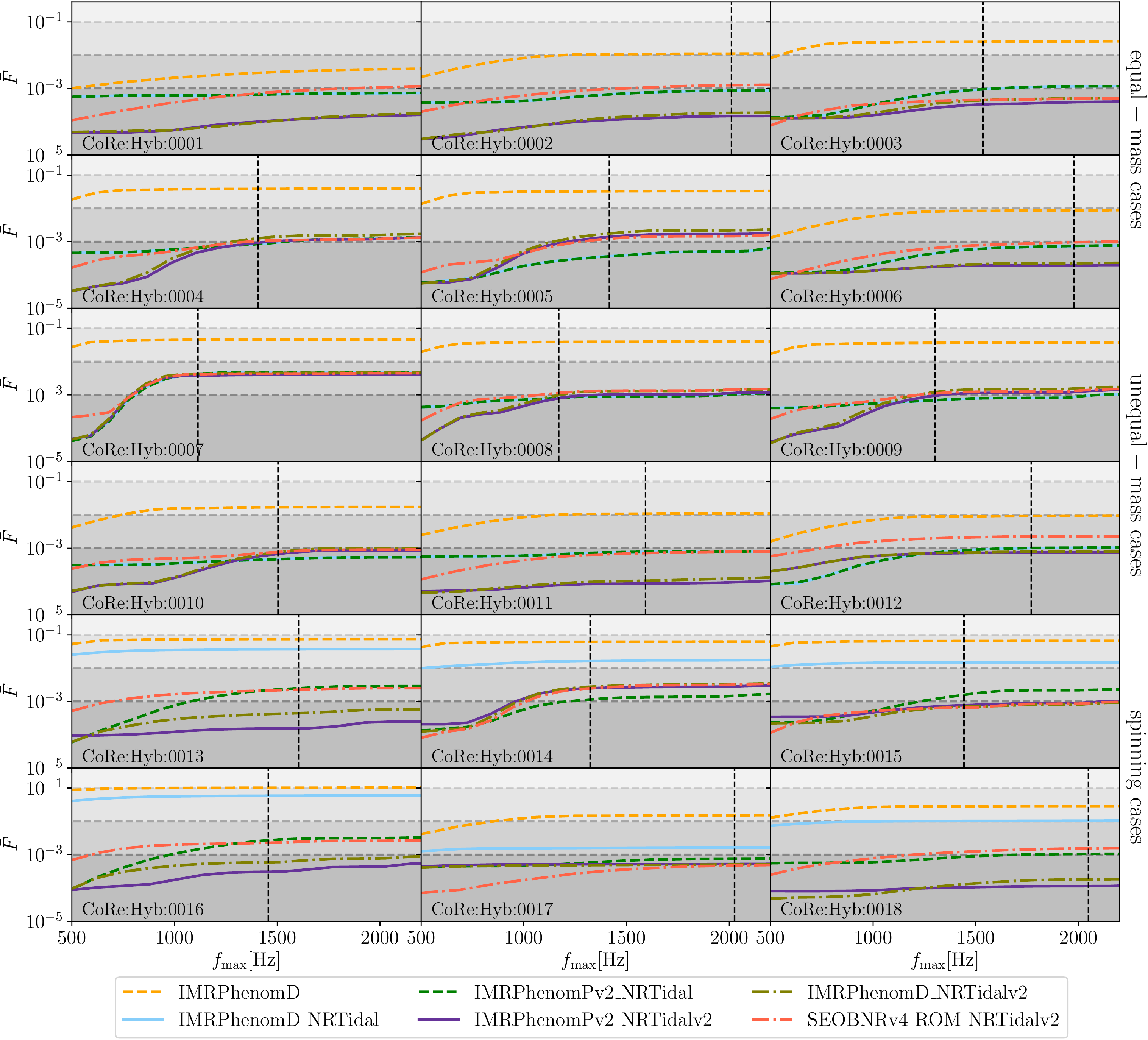}
\caption{Mismatch with respect to the \texttt{TEOBResumS}-NR hybrids.
We mark the merger frequency with a vertical dashed line. For CoRe:Hyb:0001
the merger happens at $2567\rm Hz$. 
The horizontal dashed lines mark mismatches of $10^{-3}$, $10^{-2}$, and $10^{-1}$.}
\label{fig:hybrid_matches}
\end{figure*}

To validate the new \NRTidalp model, we compare our LALSuite implementation
against a set of target waveforms combining \texttt{TEOBResumS} and NR data
by computing the mismatch.
Those waveforms have been constructed for Ref.~\cite{Dietrich:2018uni} and are publicly
available under \texttt{www.computational-relativity.org}~\cite{Dietrich:2018phi}.
We refer to~\cite{Dietrich:2018uni} for further details.
The main properties of these target waveforms are summarized 
in Table~\ref{tab:hybrids}.

\textbf{Mismatch computation:}
We compute the mismatch according to
\begin{equation}
\bar{F} = 1 - \max_{\phi_c,t_c} \frac{(h_1(\phi_c,t_c)|h_2)}{\sqrt{(h_1|h_1)(h_2|h_2)}}\,,
\label{eq:mismatch}
\end{equation}
where $\phi_c,t_c$ are an arbitrary phase and time shift.
The noise-weighted overlap is given by
\begin{equation}
 (h_1 | h_2) = 4 \Re \int_{f_{\rm min}}^{f_{\rm
 max}} \frac{\tilde{h}_1(f) \tilde{h}_2(f)}{S_n(f)} \, \text{d} f \ ,
\end{equation}
where tildes denote the Fourier transform, $S_n(f)$ is the spectral density of the detector
noise, and $f$ is the GW frequency (in the frequency domain).
We used the Advanced LIGO zero-detuning, high-power (\verb#ZERO_DET_high_P#) noise curve
of~\cite{adligo-psd} for our
analysis\footnote{We note that this noise curve has recently been updated slightly~\cite{updated_adligo-psd},
but for consistency with Ref.~\cite{Dietrich:2018uni} we employ the old noise curve.}
with a fixed $f_{\rm min} = 30 \, {\rm Hz}$ and a variable $f_{\rm max}$ ranging
from 500 Hz up to the merger frequency ($f_{\rm mrg}$) reported in Table~\ref{tab:hybrids}.

\textbf{Mismatch with respect to hybrid waveforms:}
We compute the mismatch for $18$ \texttt{TEOBResumS}-NR hybrid waveforms (Table~\ref{tab:hybrids})
against a range of different phenomenological models:
\texttt{IMRPhenomD}~\cite{Husa:2015iqa,Khan:2015jqa} (no tidal effects),
\texttt{IMRPhenomD\_NRTidal} (incorporating tidal effects
using the NRTidal model of \cite{Dietrich:2018uni} 
but no quadrupole-monopole self-spin terms),
\texttt{IMRPhenomPv2\_NRTidal} incorporating tidal effects
using the NRTidal model of \cite{Dietrich:2018uni} including 
quadrupole-monopole self-spin terms up to 3PN,
and the new model \texttt{IMRPhenomPv2\_NRTidalv2}.
In addition, we include the new \texttt{SEOBNRv4\_ROM\_NRTidalv2} and 
\texttt{IMRPhenomD\_NRTidalv2} approximants. 
We evaluate the waveform models at the parameters of the hybrids
reported in Table~\ref{tab:hybrids} with an initial frequency of 
$30 \, {\rm Hz}$.

Generally, we find that \texttt{IMRPhenomPv2\_NRTidalv2} performs 
as well or better than \texttt{IMRPhenomPv2\_NRTidal}, except for 2 cases. 
For all configurations the mismatch stays 
below $5\times 10^{-3}$ even for maximum frequencies at or above 
the merger frequency. 
In addition, our comparisons show again that the inclusion of the quadrupole-monopole
terms is important even for astrophysically reasonable spins---see~\cite{Harry:2018hke,Dietrich:2018uni,Samajdar:2019ulq} for previous studies. 
In most cases the mismatches between the hybrids and \texttt{IMRPhenomD\_NRTidalv2}
are marginally smaller compared to \texttt{SEOBNRv4\_ROM\_NRTidalv2}. 
Even less notable are the differences between \texttt{IMRPhenomD\_NRTidalv2}
and \texttt{IMRPhenomPv2\_NRTidalv2} which are dominantly driven by the 
additional 3.5PN spin-spin and cubic-in-spin contributions in 
\texttt{IMRPhenomPv2\_NRTidalv2}. The additional tidal amplitude corrections have almost 
a negligible effect; cf.~the non-spinning configurations in Fig.~\ref{fig:hybrid_matches}.

Our comparison shows that the \texttt{TEOBResumS}-NR hybrids are well described 
by the new approximant and that no additional pathologies (in the low frequency regime) 
are introduced during recalibration.  

\subsection{Cross validation against SEOBNRv4T}
\label{sec:validation:mismatches}

As a final check of the approximant, we compute 
the mismatch between the \texttt{IMRPhenomPv2\_NRTidalv2} 
and the \texttt{SEOBNRv4T} model
for a number of randomly sampled configurations.
We compare these mismatches with mismatches between \texttt{IMRPhenomPv2} and
\texttt{IMRPhenomPv2\_NRTidalv2} to give an impression of the importance of tidal effects.
The match computation is restricted to the frequency interval of $f\in[40,2048] \ \rm Hz$.
We have tested starting frequencies of $25 \ \rm Hz$ and $30 \ \rm Hz$ for a smaller number of cases
and obtained smaller mismatches than for the $40 \ \rm Hz$ initial frequency.
Therefore, to save computational costs and to provide a conservative estimate, 
we use a minimum frequency of $40 \ \rm Hz$.

\subsubsection{Non-spinning Configurations}

\begin{figure}[t]
\includegraphics[width=.475\textwidth]{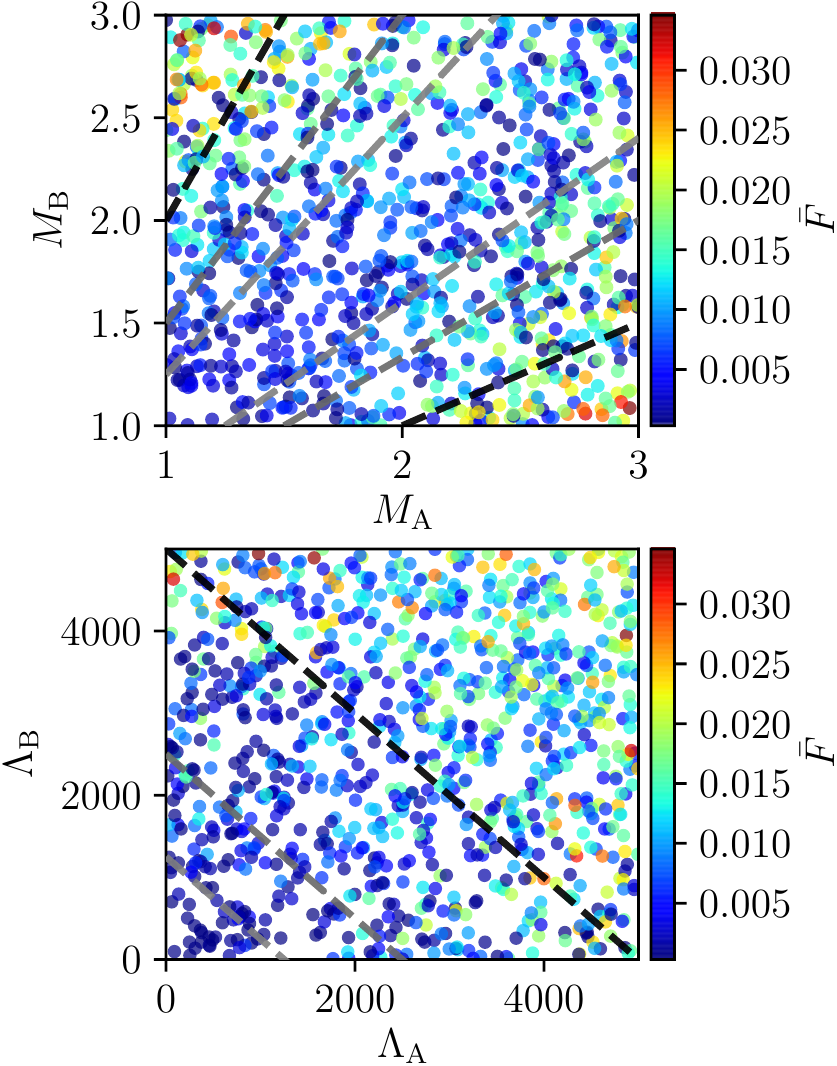}
\caption{Mismatch between \texttt{IMRPhenomPv2\_NRTidalv2} and \texttt{SEOBNRv4T}.
We consider $1000$ randomly distributed non-spinning configurations with $M_A,M_B\in[1,3] \, M_\odot$
(here we relax our usual assumption that $M_A \geq M_B$)
and $\Lambda_A,\Lambda_B \in [0,5000]$. The mismatches are computed within the
frequency interval $f\in[40,2048] \ \rm Hz$ and we use a sampling rate of $8192\ \rm Hz$.
We mark in the top panel mass ratios of $1.25;1.5;2.0$ with diagonal gray, dark gray, black dashed
lines, respectively. Similarly, $\Lambda_A+\Lambda_B=1250;2500;5000$ are marked in the bottom panel. }
\label{fig:PS_nospin}
\end{figure}

We start this analysis by considering non-spinning configurations.
For this purpose, we select $1000$ samples with flat priors
in $M_A,M_B\in[1,3] \, M_\odot$ and
$\Lambda_A,\Lambda_B \in [0,5000]$.
The final analysis is shown in Fig.~\ref{fig:PS_nospin},
where we compare the \texttt{IMRPhenomPv2\_NRTidalv2} 
and \texttt{SEOBNRv4T} approximant.
For non-spinning configurations the mismatches between
\texttt{IMRPhenomPv2\_NRTidalv2} and \texttt{SEOBNRv4T} 
are below $0.034$ for our set of configurations.
The largest difference is found for large mass ratios;
cf.~upper left and lower right corners of the top panel.
For better visualization, we mark mass ratios of $q=1.25;1.5;2.0$
by diagonal gray, dark gray, and black lines, respectively. 
Restricting to mass ratios below $1.5$, we find a largest 
mismatch of $\bar{F}=0.024$

In addition, our analysis shows that for larger tidal deformabilities
the mismatch between the two models tends to increase;
cf.~upper right corner of the right panel in Fig.~\ref{fig:PS_nospin}.
We mark in the plot $\Lambda_A+\Lambda_B=1250;2500;5000$
with gray, dark gray, and black lines. 
Restricting our analysis to $\Lambda_A+\Lambda_B<2500$
leads to a maximum mismatch of $\bar{F}=0.016$.

Overall, the average mismatch between \texttt{IMRPhenomPv2\_NRTidalv2} 
and \texttt{SEOBNRv4T} for our dataset is $0.009$. Interestingly, 
if we restrict our analysis to the more physical parameter space in which 
the more massive star has the smaller tidal deformability,\footnote{For equations of state with no phase transition,
the dimensionless tidal deformability is a monotonically decreasing function of the star's mass---see, e.g., Fig.~1 in~\cite{Chatziioannou:2015uea} for an illustration.
However, in cases with a phase transition that yield twin stars, the tidal deformability is no longer a monotonically decreasing (or even a single-valued) function of mass,
as illustrated in, e.g., Refs.~\cite{Sieniawska:2018zzj,Han:2018mtj}. Of course, even in twin star cases, the deviations from monotonic decrease and single-valuedness are not large, while the parameters we generate by the aforementioned random sampling can have significant violations.} the average 
mismatch decreases by roughly a factor of 2 to $0.0059$. 

Consequently, we find for non-spinning configurations a good 
agreement between the tidal EOB model \texttt{SEOBNRv4T}  
and \texttt{IMRPhenomPv2\_NRTidalv2}.\footnote{We note that as cross-validation 
of the implementation of the other approximants, we also tested the mismatch between
 \texttt{IMRPhenomPv2\_NRTidalv2} and \texttt{SEOBNv4\_ROM\_NRTidalv2} 
 and find an average mismatch of $\sim 5 \times 10^{-4}$.}

\subsubsection{Spinning Configurations}

\begin{figure*}[t]
\includegraphics[width=1.\textwidth]{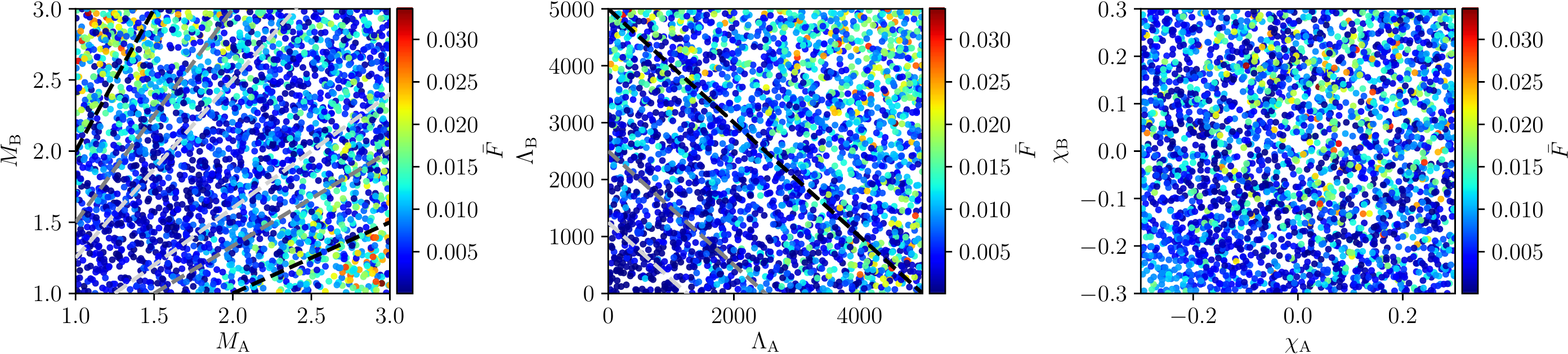}
\includegraphics[width=1.\textwidth]{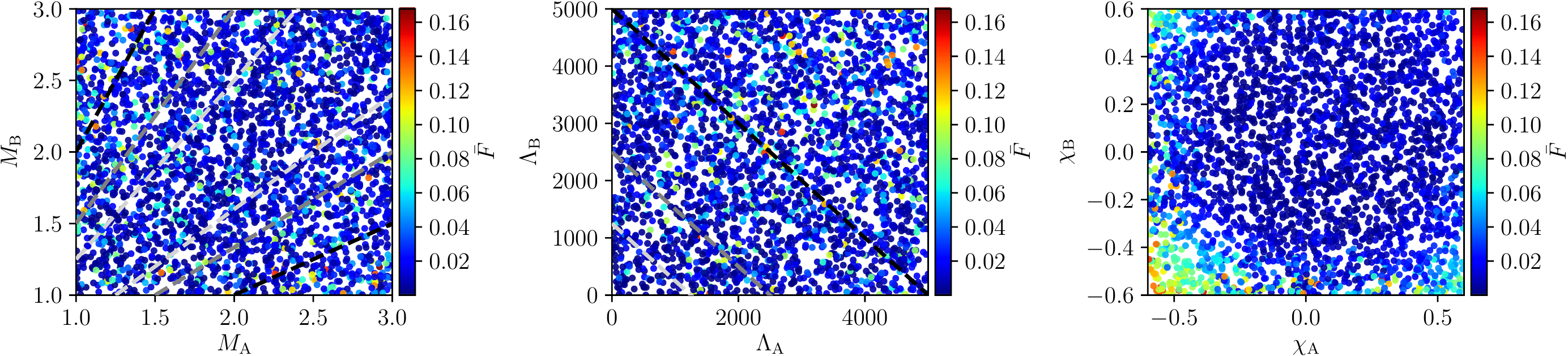}
\caption{Mismatch between \texttt{IMRPhenomPv2\_NRTidalv2} and \texttt{SEOBNRv4T}.
We consider randomly distributed configurations with $M_{A,B}\in[1,3] \, M_\odot$,
$\Lambda_{A,B} \in [0,5000]$, and $\chi_{A,B} \in [-0.30,0.30]$, as well as 
$\chi_{A,B} \in [-0.70,0.70]$.
The mismatches are computed within the
frequency interval $f\in[40,2048] \ \rm Hz$ 
and a sampling rate of $8192\ \rm Hz$ is employed.
We select $3000$ random samples for configurations with spins 
$\chi_{A,B} \in [-0.30,0.30]$ (top panels) 
and $3000$ samples for $\chi_{A,B} \in [-0.60,0.60]$ 
(bottom panels). }

\label{fig:PS_spin}
\end{figure*}

We further consider spinning configurations using flat priors
$M_{A,B}\in[1,3] \, M_\odot$,
$\Lambda_{A,B} \in [0,5000]$,
and $\chi_{A,B} \in [-0.30,0.30]$ 
as well as $\chi_{A,B} \in [-0.60,0.60]$. 
For both prior choices we select $3000$ 
randomly distributed samples. 

If we consider spins within $\chi_{A,B} \in [-0.30,0.30]$ 
(upper panels of Fig.~\ref{fig:PS_spin}), we find a maximum 
mismatch of $\bar{F}=0.034$, 
which is comparable with the non-spinning result presented before.
Overall, the average mismatch of our $3000$ samples for spins within 
$\chi_{A,B} \in [-0.30,0.30]$ is $\bar{F}=0.0072$. 
For the same set of configurations, the average mismatch with respect to 
\texttt{IMRPhenomPv2\_NRTidal} is $\bar{F}=0.0092$, 
i.e., $25\%$ larger. 
Furthermore, we find that, as for the non-spinning cases, 
the largest mismatches are obtained for configurations which 
have large mass ratios and large tidal deformabilities. 
If only spin magnitudes up to $|\chi_{A,B}|\leq 0.3$ are 
considered, we do not find a noticeable spin effect. 

However, spin effects become important for large spin magnitudes. 
For spin magnitudes up to $0.6$, the 
largest mismatches between \texttt{SEOBNRv4T} and 
\texttt{IMRPhenomPv2\_NRTidalv2} are found for large 
anti-aligned spins, i.e., the lower left corner of the right-most
bottom panel of Fig.~\ref{fig:PS_spin}.
The maximum mismatch is ${\bar{F}=0.167}$ for our randomly chosen 
set of configurations. 

Comparing average values, we find that while the average mismatch is 
$0.043$ between the original \texttt{IMRPhenomPv2\_NRTidal} model and \texttt{SEOBNRv4T}, 
the average mismatch decreases to $0.021$ between the
\texttt{IMRPhenomPv2\_NRTidalv2} model and \texttt{SEOBNRv4T}, 
i.e., much better agreement 
is found within this large region of the parameter space. 

The disagreement between \texttt{SEOBNRv4T} and 
\texttt{IMRPhenomPv2\_NRTidalv2} for large anti-aligned spins needs 
further investigation and requires additional NR simulations 
in regions of the parameter space which are currently not covered. 
Note that for the largest anti-aligned spin, high-quality NR setups 
(CoRe:BAM:0062) the NSs have only a spin of $\chi_{A,B}=-0.10$. 
For this physical configuration, both waveform approximants 
(\texttt{SEOBNRv4T} and \texttt{IMRPhenomPv2\_NRTidalv2}) describe
the data within the estimated uncertainty. 

\section{Summary}
\label{sec:summary}

In this article we have presented our most recent update of 
the \NRTidal model. The model gives a closed analytical expression 
for tidal effects during the BNS coalescence and can be added 
to an arbitrary BBH baseline approximant. 
We added the new \NRtidalp approximant to 
\texttt{IMRPhenomPv2}~\cite{Hannam:2013oca, Khan:2015jqa} 
to obtain a frequency-domain precessing BNS approximant 
as well as to the (frequency-domain) \texttt{SEOBNRv4\_ROM}~\cite{Bohe:2016gbl} and the 
\texttt{IMRPhenomD}~\cite{Khan:2015jqa} approximants to allow an improved 
and fast modeling of spin-aligned systems.

Our main improvements in comparison to the initial \NRTidal model are:
\begin{enumerate}[(i)]
 \item a recalibration of the tidal phase to improved 
 NR data incorporating additional analytical knowledge 
 for the low frequency limit;
 \item the addition of a tidal amplitude correction to the model;
 \item incorporation of higher order (3.5PN) quadrupole and 
 octupole information to the spin sector of the model. 
\end{enumerate}

We also hope to further improve the \NRtidalp 
model for higher mass ratios to allow 
an accurate description of high mass ratio systems. 
Such an extension requires additional high-quality
NR simulations for a variety of 
different mass ratios. 

An additional improvement would be the incorporation of the effect of 
$f$-mode resonances as recently computed in Refs.~\cite{Andersson:2019dwg,Schmidt:2019wrl}, 
the incorporation of an updated precession dynamics 
as used in~\cite{Khan:2018fmp}, or the incorporation of 
higher modes~\cite{London:2017bcn,Cotesta:2018fcv}.

We have compared the \texttt{IMRPhenomPv2\_NRTidalv2} model 
with high resolution numerical relativity data and found agreement within the estimated 
uncertainty for all NR data with clear convergence. Overall, the performance of \texttt{IMRPhenomPv2\_NRTidalv2} 
is comparable with state-of-the-art tidal EOB models. 

This accuracy was verified by the mismatch computation between  
\texttt{IMRPhenomPv2\_NRTidalv2}  and \texttt{TEOBResumS}-NR hybrid waveforms, 
for which mismatches are well below $5\times10^{-3}$. 

We concluded the performance test of the model with a mismatch computation 
with respect to the tidal EOB model \texttt{SEOBNRv4T}. 
For non-spinning cases (or cases with small spins as employed in the 
low spin prior of the LVC analysis) the mismatch computed from a 
starting frequency of $40\ \rm Hz$ never exceeds $\bar{F}\approx 0.034$
for $M_{A,B}\in [1,3]\, M_\odot$ and $\Lambda_{A,B}\in [0,5000]$. 
Considering spinning setups ($\chi_{A,B}\in [-0.6,0.6]$), 
the mismatch increases to a maximum of $\bar{F}=0.164$. 

\begin{acknowledgments}

  We acknowledge fruitful discussions with 
  Sebastiano Bernuzzi, Tanja Hinderer, Alessandro Nagar, 
  Patricia Schmidt, Ka Wa Tsang, and Chris Van Den Broeck 
  and thank Alessandra Buonanno, Michael P\"urrer, and Ulrich Sperhake 
  for comments on the manuscript. 
  We thank Sarp Akcay and Nestor Ortiz for the BAM:0001 high resolution data. 
  
  T.~D.~acknowledges support by the European Union's Horizon
  2020 research and innovation program under grant
  agreement No 749145, BNSmergers. 
  A.~S.~and T.~D.~are supported by the research programme
  of the Netherlands Organisation for Scientific Research (NWO).
  S.~K.~acknowledges support from the Max Planck Society’s Independent
  Research Group Grant.
  N.~K.~J.-M.\ acknowledges support
  from STFC Consolidator Grant No.~ST/L000636/1. Also, this work has received funding from the
  European Union's Horizon 2020 research and innovation programme under the Marie
  Sk{\l}odowska-Curie Grant Agreement No.~690904.
  R.~D.~acknowledges support from DFG grant BR 2176/5-1.
  W.~T.\ was supported by the National Science Foundation
  under grant PHY-1707227.
  
  The computation of the numerical relativity waveform was performed
  on the Minerva cluster of the Max-Planck
  Institute for Gravitational Physics.
\end{acknowledgments}

\appendix

\section{Alternative formulation of a tidal amplitude correction}
\label{app:tidal_amp}

\subsection{Tidal amplitude corrections in the time domain}

As in the frequency domain, the BNS waveform time
domain amplitude
can be obtained by augmenting existing BBH-models
with additional tidal corrections $A_{\rm T}$, i.e.,
\begin{equation}
A_{\rm BNS}= A_{\rm BBH} + A_{\rm T}.
\end{equation}
Refs.~\cite{Damour:2012yf,Banihashemi:2018xfb}
present the tidal amplitude corrections
for the leading and next-to-leading order
\begin{small}
\begin{equation}
\begin{split}
 \label{eq:AmpT_PN}
 A_{\rm T} &=  \frac{8 M \nu }{D_L} x \sqrt{\frac{\pi}{5}} \biggl\{ \Lambda_A X_A^4 x^5\biggl[ 3 (1 + 2 X_B) \\
 &\quad +  \frac{63-15X_B-205 X_B^2-45X_B^3}{14} x  + \mathcal{O}(x^{3/2})\biggr]\\
 &\quad + [A \leftrightarrow B] \biggr\}.
\end{split}
\end{equation}
\end{small}

While Eq.~\eqref{eq:AmpT_PN} describes the tidal amplitude corrections
for small frequencies, it loses validity close to the moment of merger.
For extreme cases, i.e., stiff EOSs and low NS masses,
the additional amplitude corrections can become larger than $A_{\rm BBH}$
causing the overall amplitude to be negative.
Thus, a further calibration to NR or EOB data is required.
We will employ the quasi-universal relations, which allow an
EOS-independent description of important quantities at the moment of merger
(merger frequency, merger amplitude, reduced binding energy, specific orbital
angular momentum, and GW luminosity)~\cite{Bernuzzi:2014kca,Takami:2014tva,
dbt_mods_00029293,Zappa:2017xba}.
As shown in Fig.~6.7 of Ref.~\cite{dbt_mods_00029293}, the GW amplitude
at the merger follows a quasi-universal relation as a function of
the tidal coupling constant
\begin{equation}
\kappa^T_{2} = 2 \left[
\frac{X_B}{X_A} \left(\frac{X_A}{C_A}\right)^5 k^A_2 +
\frac{X_A}{X_B} \left(\frac{X_B}{C_B}\right)^5 k^B_2 \right],
\label{eq:kappa2}
\end{equation}
namely,
\begin{small}
\begin{align}
  & D_L A^{\rm mrg}/(\nu M )  = \nonumber \\ & 1.6498 \frac{1+2.5603\cdot
  10^{-2}\kappa_2^T-1.024\cdot 10^{-5}(\kappa_2^T)^2}
  {1+4.7278\cdot10^{-2}\kappa^T_2}.
\label{eq:Amrg_qu}
  \end{align}
\end{small}

We note that a straightforward extension of Eq.~\eqref{eq:Amrg_qu}
[Eq.~(6.15d) of Ref.~\cite{dbt_mods_00029293}],
would be the incorporation of a larger number of NR
simulations as publicly available under
\texttt{www.computational-relativity.org}, Ref.~\cite{Dietrich:2018phi}.
However, we postpone this to future work~\cite{Zappa_inprep},
in which a more general discussion about quasi-universal relations
during the BNS coalescence will be given.

To incorporate Eq.~\eqref{eq:Amrg_qu} in Eq.~\eqref{eq:AmpT_PN},
we extend the analytical knowledge with an additional, unknown higher order PN-term
and define the \NRTidal amplitude correction as
\begin{small}
\begin{equation}
\label{eq:td_tidal_amp}
 A_{\rm T}^{\rm NRTidal} = \frac{8  M  \nu}{D_L} x \sqrt{\frac{\pi}{5}}\left(\hat{c}_A \kappa_A x^5 \frac{1+ \hat{c}_1^A x}{1+ \hat{d} x }
 + \hat{c}_B \kappa_B x^5 \frac{1 + \hat{c}_1^B x }{1+ \hat{d} x}\right),
\end{equation}
\end{small}
where the individual terms $\hat{c}_A, \hat{c}_1^A, \hat{c}_B, \hat{c}_1^B$ 
can be obtianed from Eq.~\eqref{eq:AmpT_PN} once we express $\Lambda_{A,B}$
in terms of $\kappa_{A,B}$.

Enforcing
\begin{small}
\begin{equation}
 A^{\rm mrg} (x^{\rm mrg}) =
 A_{\rm BBH}(x^{\rm mrg}) +A_{\rm T}^{\rm NRTidal} (x^{\rm mrg})
\end{equation}
\end{small}
gives us the unknown parameter $d$ according to
\begin{small}
\begin{align}
 d & = \left\{ \frac{8 M \nu  x^5}{D_L \ \Delta A} \sqrt{\frac{\pi}{5}} \left[\hat{c}_A \kappa _A \left(1+ \hat{c}_1^A x \right) \right. \right. \nonumber \\
   & \quad  + \left. \left. \hat{c}_B \kappa _B \left(1+ \hat{c}_1^B x\right)\right]- \frac{1}{x} \right\}_{x=x^{\rm mrg}}
\end{align}
\end{small}
with $\Delta A = A^{\rm mrg} (x^{\rm mrg})- A_{\rm BBH}(x^{\rm mrg})$.

We note that although the outlined approach has been tested
for a selected number of cases, we did not implemented it in
LALSuite due to the large computational costs inherent to time
domain waveform approximants.

\subsection{Frequency domain amplitude corrections by SPA}

In addition to the frequency domain amplitude correction
presented in the main text, we also want to present a possible
alternative way to augment the frequency domain binary black hole
amplitude with tidal correction.
For this purpose we use the SPA to
obtain
the frequency-domain amplitude from Eq.~\eqref{eq:td_tidal_amp}.
Following the SPA approach,
\begin{equation}
\tilde{A} = \frac{1}{2} \sqrt{\frac{2 \pi}{\ddot{\phi}}} A, \label{eq:tAT_SPA}
\end{equation}
where $\ddot{\phi}$ refers to the second time derivative of $\phi$.

Using Eq.~\eqref{eq:SPA} [see also Eq.~(14) in~\cite{Damour:2012yf}; note that the published version is missing an equals sign], Eq.~\eqref{eq:tAT_SPA} can be rewritten as
\begin{equation}
\tilde{A} = A \sqrt{\frac{\pi}{2}} \ \sqrt{\frac{\text{d}^2 \psi}{\text{d} \mo^2}}.
\end{equation}
Inserting $A=A_{\rm BBH}+A_{\rm T}$ and $\psi = \psi_{\rm BBH}+ \psi_{T}$ leads to
\begin{equation}
\tilde{A} =
(A_{\rm BBH}+A_{\rm T}) \sqrt{\frac{\pi}{2}} \ \sqrt{\frac{\text{d}^2 \psi_{\rm BBH}}{\text{d} \mo^2}+
\frac{\text{d}^2 \psi_{\rm T}}{\text{d} \mo^2}}.
\end{equation}
Treating the tidal phase correction as a small change of the underlying BBH waveform,
we rewrite the expression as
\begin{equation}
\tilde{A} =
(A_{\rm BBH}+A_{\rm T}) \sqrt{\frac{\pi}{2}} \  \sqrt{\frac{\text{d}^2 \psi_{\rm BBH}}{\text{d} \mo^2}}
\sqrt{1+\underbrace{\frac{\text{d}^2 \psi_{\rm T}}{\text{d} \mo^2} \bigg/ \frac{\text{d}^2 \psi_{\rm BBH}}{\text{d} \mo^2}}_\epsilon}.
\end{equation}
Linearizing in $\epsilon = \mathcal{O}(\kappa)$ and neglecting terms proportional to $\kappa^2$ [noting that $A_{\rm T} = \mathcal{O}(\kappa)$] leads to
\begin{equation}
\tilde{A} =
\underbrace{A_{\rm BBH}\sqrt{\frac{\pi}{2}} \  \sqrt{\frac{\text{d}^2 \psi_{\rm BBH}}{\text{d} \mo^2}}}_{\tilde{A}_{\rm BBH}} \left(1+\frac{\epsilon}{2}\right)+
A_{\rm T} \underbrace{\sqrt{\frac{\pi}{2}} \  \sqrt{\frac{\text{d}^2 \psi_{\rm BBH}}{\text{d} \mo^2}}}_{\tilde{A}_{\rm BBH}/A_{\rm BBH}}.
\end{equation}
Thus, the final expression is given as
\begin{equation}
 \frac{\tilde{A}}{\tilde{A}_{\rm BBH}} = 1 +\frac{A_{\rm T}}{A_{\rm BBH}}
 +\frac{1}{2} \frac{\text{d}^2 \psi_{\rm T}}{\text{d} \mo^2}\bigg/ \frac{\text{d}^2 \psi_{\rm BBH}}{\text{d} \mo^2}.
\label{eq:FD_A_SPA}
\end{equation}
The approach outlined in this appendix, i.e., Eq.~\eqref{eq:FD_A_SPA}, leads to
much larger computational cost than the Pad{\'e} approximant [Eq.~\eqref{eq:A_FD_NRTP}],
which is why we chose the easier and more straightforward implementation shown in the
main body of the paper.
However, this additional approach might become relevant for a potential improvement/extension
in the future.

\section{Extension of the NRTidal phase incorporating additional mass ratio dependence}
\label{app:mass_ratio}

We now outline a possible extension 
of the \NRTidalp model which incorporates 
additional analytically known mass-ratio dependence. 
Since we do not find such an extension to perform better 
in our tests and to reduce computational costs, we limited the mass ratio dependence 
in the tidal phase simply to the prefactor $\propto  \kappa_{\rm eff}^T/\nu$ in the current
implementation of the model. 

However, it is possible to recast Eq.~\eqref{eq:NRTidal_FD}
as
\begin{equation}
 \psi_{\rm T} (x) = - \kappa_A c^A_0 x^{5/2} \tilde{P}_{\rm NRTidal}^A(x) 
                    - \kappa_B c^B_0 x^{5/2} \tilde{P}_{\rm NRTidal}^B(x)
\end{equation}
with $c_0^A = -\frac{3}{16 X_A X_B} \left(12 + \frac{X_{A}}{X_{B}}\right)$ (and similarly for $c_0^B$). 

The individual Pad\'e approximants $\tilde{P}_{\rm NRTidal}^A(x)$, $\tilde{P}_{\rm NRTidal}^B(x)$ 
are similar to Eq.~\eqref{eq:Pade_FD} together with the constraints in
Eqs.~\eqref{eq:Pade_FD_constraints}, 
but the known PN coefficients $\tilde{c}_1^{A,B},
\tilde{c}_{3/2}^{A,B},\tilde{c}_2^{A,B}, \tilde{c}_{5/2}^{A,B}$
have a mass ratio dependence as given in Ref.~\cite{Damour:2012yf}. 

Due to the limited set of high-quality NR data, 
the fitting coefficients in
Eqs.~\eqref{eq:Pade_FD_coeffs}
($\tilde{d}_1,\tilde{d}_2,\tilde{n}_{5/2},\tilde{n}_3$)
can only be determined for the equal mass case. 

We find that while such a choice of the coefficients leads to a correct mass ratio 
dependence for the low frequency limit, the higher frequency phase is described worse compared 
to the \NRTidalp approximation given in the main text. We suggest that this is caused by 
the inconsistency introduced by adding the mass-ratio dependence in only some of the Pad\'e 
coefficients. 

\bibliography{refs}

\end{document}